\documentclass[aps,reprint,pra,longbibliography,superscriptaddress]{revtex4-1}
\usepackage{amssymb}
\usepackage{amsmath}
\usepackage{dcolumn}
\usepackage{bm}
\usepackage{graphicx}
\usepackage{mathrsfs}
\usepackage[colorlinks,linkcolor=blue,anchorcolor=blue,citecolor=blue,urlcolor=blue]{hyperref}

\begin{document}

\title{Interfacing a topological qubit with a spin qubit in a hybrid quantum system}% Force line breaks with \\

\author{Bo Li}
\affiliation{Shaanxi Province Key Laboratory of Quantum Information and Quantum Optoelectronic Devices,
             Department of Applied Physics, Xi'an Jiaotong University, Xi'an 710049, China}%
\affiliation{Department of Physics and Astronomy, Purdue University, West Lafayette, IN 47907, USA}
\author{Peng-Bo Li }
\email{lipengbo@mail.xjtu.edu.cn}
\author{Yuan Zhou }
\author{Jie Liu }
\author{Hong-Rong Li }
\author{Fu-Li Li }
\affiliation{Shaanxi Province Key Laboratory of Quantum Information and Quantum Optoelectronic Devices,
             Department of Applied Physics, Xi'an Jiaotong University, Xi'an 710049, China}%

\date{\today}% It is always \today, today,
             %  but any date may be explicitly specified

\begin{abstract}
We present and analyze a hybrid quantum system that interfaces a Majorana-hosted semiconductor nanowire
with a single nitrogen-vacancy (NV) center via a magnetized torsional cantilever.
We show that, the torsional mode of the mechanical resonator can strongly couple to the Majorana qubit and the spin qubit
simultaneously, which allows to interface them for quantum state conversion through a dark state protocol.
This work provides a promising interface between the topological qubit and the conventional  spin qubit, and can find
interesting applications in quantum information and computation.
\end{abstract}

%\pacs{Valid PACS appear here}% PACS, the Physics and Astronomy
                             % Classification Scheme.
%\keywords{Suggested keywords}%Use showkeys class option if keyword
                              %display desired
\maketitle

\section{introduction}

Majorana Fermion quasiparticles (MFs) realized in condensed matter systems  have attracted great interest
in recent years \cite{RMP-87-137,NR-3-52,Natrue-74-556,NP-5-614,PRB-97-205404}, because they provide a very intriguing platform for quantum science and technology. The proposed realizations include $p_{x}+ip_{y}$ superconductors \cite{PRB-61-10267},
edges of 2D topological insulators \cite{PRL-100-096407}, quantum Hall states at filling factor 5/2 \cite{RevModPhys-80-1083},
and 1D semiconducting quantum wires with strong spin-orbit interactions \cite{PRL-104-040502,PRB-81-125318,PRL-105-077001,PRA-90-012323,PRL-105-177002}.
In these systems, the spin-orbit coupled semiconductor wires with proximity-induced superconductivity are of particular interests
for their simple architectures. For the 1D quantum nanowire,
the MFs are predicated to localize through the  competition between the superconducting proximity effect and
the Zeeman splitting. Recently, several experiments aiming at establishing the existence of MFs have been
reported \cite{Science-336-1003,NL-12-6414,NatPhys-8-887,NatPhys-8-795}.
Meanwhile, it has been shown that the Majorana qubits can be controlled by
integrating with the conventional ones, such as the flux qubits and the spin qubits \cite{NJP-12-125002,PRA-79-040303,PRL-106-090503,
PRL-107-210502,PhysRevB-94-045316,PhysRevB-94-174514}.

Topological qubits are one of the most promising candidates for quantum computation due to their extreme robustness against local fluctuations.
However, the technical challenges are considerable:  (\romannumeral1). it is well known that braiding operations of MFs cannot
generate a complete set of universal quantum logic gates required for quantum computation;
(\romannumeral2). it is still challenging to operate and readout the quantum states of the MFs.
To overcome these problems, several types of hybrid quantum architectures coupling the topological qubits
to the conventional ones have been proposed.
One potential route is  controlling the  microscopic wave function of the superconductors,
which allows the strong coupling between a topological qubit and a flux qubit, or a mechanical oscillator
\cite{PRA-87-032327,PRA-88-024303,PRA-87-032339}.
Other promising schemes include utilizing the Aharonov-Casher effect
\cite{PRL-106-130504,PRL-106-130505,NP-7-450}, or  the fractional Josephson effect \cite{NJP-16-015009,PRB-88-195415,SR-5-12233}.
So far, for all of these protocols, further experimental demonstrations are needed to verify their feasibility.

For the conventional qubits,  nitrogen-vacancy (NV) centers in diamond are more attractive due to their unique features,
such as the fast microwave manipulation, optical preparation and detection, and long coherence time even at room temperature \cite{PR-528-1,NL-12-3920,NC-4-1743}.
In recent years, there has been considerable effort to integrate NV centers in hybrid quantum systems,
 allowing the preparation of fantastic quantum states \cite{prb-94-205118,PRA-83-022302,SR-7-14116,pra-96-062333,PRA-88-033614,PRA-83-054305},
design of quantum logic gates \cite{Natrue-514-72,PRL-109-070502,PRB-96-205149,prb-96-245418,pra-97-062318}, and storage
or transfer of quantum information \cite{pra-91-042307,PRA-84-010301,pra-88-012329,PRA-96-032342,PRB-87-144516,PRB-90-195112}.
To further explore the potential of hybrid quantum systems, it will be very appealing to
interface topological qubits with NV centers coherently.
Integrating the topological degrees of freedom with the complete gate operations of NV spins,
will provide a promising multi-qubit platform for quantum computation.
However, the energy mismatch between these two kinds of qubits possesses a major obstacle for their coupling.

In this paper, we consider a hybrid quantum device interfacing a topological qubit with a single NV center via a magnetized torsional cantilever.
The topological qubit under consideration is realized in a spin-orbit coupled semiconductor nanowire, which is
embedded on top of an $s$-wave superconductor to form a semiconductor-superconductor heterostructure.
Here the torsional cantilever is realized by an individual single-walled nanotube (SWNT) with a nano-size magnet attached.
This nanoscale mechanical structure enables us to combine an isolated NV center with the strong spin-torsion couplings.
Together with the strong topology-torsion couplings induced by the Magneto-Josephson effect \cite{PRL-112-106402},
the mechanical torsional mode can  strongly couple to both the Majorana qubit and  the  NV center.
This enables coherent quantum states conversion between the Majorana qubit and the spin qubit via a dark-state protocol.
Owing to the topologically protected robustness and the universal quantum logic gates of NV spins,
there will be more potential applications utilizing this hybrid architecture.

\section{Description of the device}

As shown schematically in Fig.~1, the hybrid quantum
device under consideration consists of a topological qubit realized in a Majorana-hosted semiconductor
nanowire, a magnetized torsional cantilever, and a single NV center.
The MFs in this architecture are realized in a spin-orbit coupled semiconductor
nanowire with proximity-induced superconductivity.
This 1D quantum nanowire could be driven to the topological phase regime, and the MF exists as a quasiparticle at the boundary
between the topological (T) and non-topological (N) regions \cite{PRB-81-125318,PRL-104-040502,PRL-105-077001,PRA-90-012323}.
Here we divide the semiconductor nanowire into three  sections to form a TNT junction \cite{PRL-105-077001,SR-5-12233},
and define the topologically protected MFs as  $\hat{\gamma}_{1/2/3/4}$.

\begin{figure}[b]
\includegraphics[width=7.4cm]{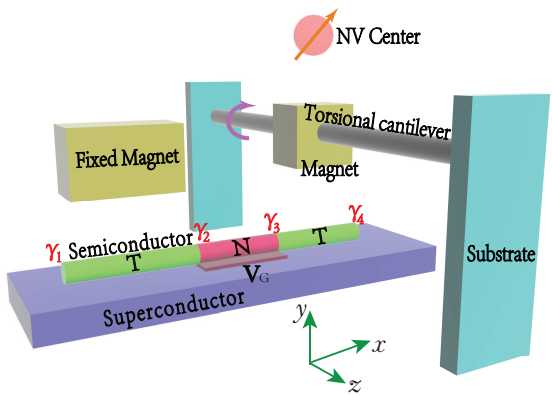}
\caption{\label{fig_1}(Color online) Schematic of a hybrid  quantum  device interfacing
a Majorana qubit and a single NV center via a  magnetized torsional cantilever.
The spin-orbit coupled semiconductor nanowire is  embedded on top of an $s$-wave superconductor,
and is divided into three regions to form a TNT junction.
The MFs $\hat{\gamma}_{1/2/3/4}$ appear at the boundaries between the T and N regions.
And for the torsional cantilever, the individual SWNT works as
a torsional spring for the suspended nanomagnet.  }
\end{figure}

In this setup, the magnetized torsional cantilever is realized by a nanomagnet with dimensions $(l,w,t)$
suspended on a SWNT. Specifically, the individual SWNT works as a torsional
spring and mechanical support, which allows the suspended nanomagnet to rotate along the tube ($z$) axis  (see Fig.~1).
Without loss of generality, we assume that  the nanomagnet can only rotate along the $z$ axis with a small amplitude,
then the  Hamiltonian of the  torsional cantilever reads
$H_{m} =\frac{1}{2}I_{z}\omega_{m}^{2}\hat{\theta}^{2}+\frac{\hat{L}^{2}}{2I_{z}}$, where $\omega_{m}$, $I_{z}$, $\hat{L}$, and $\hat{\theta}$
are the  torsional frequency, moment of inertia, angular momentum, and angular displacement, respectively.
Similar to the canonical conjugate observable $\hat{X}$ and $\hat{P}$ of the center of mass mode,
the rotational degree of freedom can be quantized \cite{PRA-73-052104}, and  the motion of the torsional cantilever can be described by
the annihilation and creation operators
\begin{equation}
\hat{b}=\frac{1}{2}(\frac{\hat{\theta}}{\theta _{zpf}}+\frac{i\hat{L}}{L_{zpf}}), \hat{b}^{\dagger }=\frac{1}{2}(\frac{\hat{\theta}}{\theta _{zpf}}-\frac{i\hat{L}}{L_{zpf}}),
\end{equation}
where $L_{zpf}=\sqrt{\hslash I_{z}\omega _{m}/2}$ and $\theta _{zpf}=\sqrt{\hslash /(2I_{z}\omega _{m})}$ denote the
zero point fluctuations of the torsional mode.

\subsection{MFs in 1D semiconductor nanowire}

We first describe the 1D quantum nanowire hosting MFs. As shown in Fig.~1,
the semiconductor nanowire is embedded on the top of an $s$-wave superconductor,
and an external magnetic field $\vec{b}=b\vec{e}_{z}$ along the $z$ axis is applied.
There are two nanomagnets  positioned straight above the nanowire:
one is stationary in the left region, and the other is free to vibrate in the right region.
We assume that the magnetic fields induced by the two magnets
is in the $xy$ plane, i.e., $\vec{B}=B\cos \theta \vec{e}_{x}+B\sin \theta \vec{e}_{y}$.
We further use the angular displacement of the torsional cantilever to denote
the relative field angle,
which implies $\theta =\theta _{r}-\theta _{l}$.

The single-particle effective Hamiltonian for  the semiconductor wire reads
$H_{0}=\hat{p}^{2}/2m^{\ast }-\mu +b\hat{\sigma} _{z}-\alpha (\hat{\vec{\sigma}}\times \hat{\vec{p}})\cdot \hat{z}$ \cite{PRL-104-040502},
where $m^{\ast }$, $\mu $, $b$, and $\alpha $  are the effective mass,  chemical potential, longitudinal magnetic field,
and Rashba spin-orbit coupling strength, respectively.
As a result of the proximity effect, the Cooper pairs tunnelling  into the nanowire can be described by the Hamiltonian
$H_{SC}=\Delta e^{i\phi }\hat{\psi} _{\uparrow }^{\dagger }(x)\hat{\psi} _{\downarrow}^{\dagger }(x)+H.c.$, with $\Delta e^{i\phi }$
%$H_{SC}=\Delta e^{i\phi }\psi _{k\uparrow }^{\dagger }(x)\psi _{-k\downarrow}^{\dagger }(x)+H.c.$, with $\Delta e^{i\phi }$
the proximity-induced gap.
We now introduce the Nambu spinor basis
$\hat{\Psi}^{T}=(\hat{\psi}_{\uparrow },\hat{\psi}_{\downarrow },\hat{\psi}%
_{\downarrow }^{\dagger },-\hat{\psi}_{\uparrow }^{\dagger })$,
and model the 1D semiconductor nanowire
with a Bogoliubov-de Gennes Hamiltonian
\begin{eqnarray}
H &=&\frac{\hslash ^{2}k^{2}}{2m^{\ast }}\hat{\tau}^{z}+\alpha k\hat{\tau}%
^{z}\hat{\sigma}^{z}-\mu \hat{\tau}^{z}+\Delta (\cos \phi \hat{\tau}%
^{x}-\sin \phi \hat{\tau}^{y})  \nonumber \\
&&-b\hat{\sigma}^{z}+B(\cos \theta \hat{\sigma}^{x}-\sin \theta \hat{\sigma}%
^{y}),
\end{eqnarray}
where the Pauli matrices $\hat{\sigma}^{i}$ and $\hat{\tau}^{i}$ represent the spin and particle-hole sectors, respectively.

As  shown in Ref. \cite{PRL-112-106402}, this semiconductor nanowire can be driven
to the topological   phase regime when $\Delta ^{2}-b^{2}<B^{2}-\mu ^{2}$, and to the  non-topological  phase
regime when $\Delta ^{2}-b^{2}>B^{2}-\mu ^{2}$ (considering only $\Delta ^{2}>b^{2}$).
Note that Refs. \cite{PRB-87-075438,PRL-107-236401} give the similar results.
In particular, the Bogoliubov operators of the quantum nanowire take a general form as
$\hat{\gamma}_{k}=u_{k\uparrow }\hat{\psi}_{k\uparrow }+u_{k\downarrow }\hat{%
\psi}_{k\downarrow }+v_{k\uparrow }\hat{\psi}_{-k\uparrow }^{\dagger
}+v_{k\uparrow }\hat{\psi}_{-k\downarrow }^{\dagger }$. For Majorana operators that satisfy
$\hat{\gamma}_{k}^{\dagger }=\hat{\gamma}_{k}$, they may only occur in the
momentum inversion symmetric condition $ k\equiv -k$, i.e., $k=0$. In the quantum nanowire, this   indicates
a phase boundary between the T  and N regions \cite{JKPS-62-1558}.

The exactly zero-energy Majorana modes  exist only  in the ideal case where the two MFs possess an infinite distance.
For a finite nanowire, the MFs gain a small energy as a result of  the weak overlap of their wave functions.
However,  the wave functions are still well localized at the boundaries, and we still have
$\hat{\gamma}^{\dagger }\simeq \hat{\gamma}$ in general. Therefore,
these MFs with nonzero energies can still be utilized to encode
quantum information. We now denote the lengths of the three nanowire sections  as $l_{l/m/r}$, then
the MFs in the 1D semiconductor nanowire can be described by an effective low energy Hamiltonian
\begin{equation}
\hat{H}_{TP}=iE_{l}\hat{\gamma}_{1}\hat{\gamma}_{2}+iE_{m}(\theta )\hat{%
\gamma}_{2}\hat{\gamma}_{3}+iE_{r}\hat{\gamma}_{3}\hat{\gamma}_{4},
\end{equation}
where $E_{l}$, $E_{m}(\theta )$, and $E_{r}$ are the coupling energies between the adjacent MFs.

\subsection{Magneto-Josephson effect}

To perform quantum information processing with the topological degree of freedom, different ways
 to manipulate the MFs have been studied \cite{NP-7-412,PRX-5-041038,PRB-82-174409,PRA-86-035602,PRB-87-060504}.
One potential route is to utilize the fraction Josephone effect \cite{PRB-87-104509,EPJB-37-349}.
Different from the conventional Josephson effect that allows tunneling by Cooper pairs only,
the fraction Josephone effect supports single electron tunneling \cite{SR-5-12233},
and possess a 4$\pi$-periodic coupling energy \cite{PRB-97-035311}.
Moreover, it has been shown that the Magneto-Josephson effect can be used to realize the
coherent coupling between the Majorana qubit and the magnetized torsional cantilever \cite{PRL-112-106402}.
In particular, the spin current induced by the Magneto-Josephson effect
serves as a mechanical torque acting on the nanomagnet, which leads to the strong topology-torsion coupling.

For the TNT junction realized in the  semiconductor nanowire, both the superconducting phases
and the magnetic fields have contributions to the fractional Josephson effects.
These contributions can be analysised by the magnetism-superconductivity duality:
if we interchange the magnetic terms $\{b,B,\theta ,\hat{\sigma}^{i}\}$ with the
superconducting terms $\{\mu ,\Delta ,\phi ,\hat{\tau}^{i}\}$, the Hamiltonian
(2) takes the same form \cite{PRB-87-075438}. In a general case, the coupling interaction
between $\hat{\gamma}_{2}$ and $\hat{\gamma}_{3}$ takes form as
$E_{M}\propto E_{M}^{0}\cos (\frac{\theta _{l}-\theta _{r}}{2})\cos (\frac{%
\phi _{l}-\phi _{r}}{2})$ \cite{JKPS-62-1558}.
In our system, we set the superconducting phases as the constants (i.e., $\phi _{l}=\phi _{m}=\phi _{r}$),
then the coupling energy $E_{m}$ can be tuned by changing the relative field angle $\theta $.

To study the Magneto-Josephson effect in the TNT junction, we now map the Bogoliubov-de Gennes Hamiltonian (2)
to the tight-binding model
\begin{eqnarray}
&&\hat{H}=-t\sum\limits_{i,\sigma }\hat{c}_{i\sigma }^{\dagger }\hat{c}%
_{i+1\sigma }+\sum\limits_{i,\sigma }(-2t+\mu _{i}+b_{i\sigma })\hat{c}%
_{i\sigma }^{\dagger }\hat{c}_{i\sigma }  \nonumber \\
&&\text{ \ \ \ }+\alpha \sum\limits_{i}(\hat{c}_{i,\uparrow }^{\dagger }\hat{%
c}_{i+1,\downarrow }-\hat{c}_{i,\downarrow }^{\dagger }\hat{c}_{i+1,\uparrow
}+H.c.) \\
&&\text{ \ \ \ }+\sum\limits_{i}(\Delta e^{i\phi }\hat{c}_{i,\uparrow
}^{\dagger }\hat{c}_{i,\downarrow }^{\dagger
}+H.c.)+\sum\limits_{i}(Be^{i\theta }\hat{c}_{i\uparrow }^{\dagger }\hat{c}%
_{i,\downarrow }+H.c.).  \nonumber
\end{eqnarray}
For this lattice version, $t=\hbar ^{2}/(2m^{\ast }a^{2})$ is the nearest-neighbor hopping strength,
$a$ is the lattice spacing constant,
 $\alpha $ is the spin-orbit coupling strength,
 %$\mu_{i\uparrow /\downarrow }=\mu _{i}\pm h$ is the  chemical potential,
 $b_{i\uparrow /\downarrow }=\pm b$ is the Zeeman splitting,
 and $\hat{c}_{i\sigma }(\hat{c}_{i\sigma
}^{\dagger })$ is the annihilation  (creation) operator of an electron with spin
$\sigma =\uparrow  (\downarrow )$.

We conduct numerical simulations by considering a 1D lattice model with 500
grid sites (300, 20, and 180 sites for the left, middle, and right regions, respectively).
Here the energy eigenvalues of the system are denoted as $\pm \varepsilon _{i}$,
with the corresponding eigenstates $|\phi _{\pm \varepsilon _{i}}\rangle$.
As shown in Figs.~2(a) and (b), the eigenstates $|\phi _{\pm \varepsilon _{1}}\rangle$
correspond to the excitations of $\hat{\gamma}_{1}$ and  $\hat{\gamma}_{4}$, while the eigenstates $|\phi _{\pm \varepsilon _{2}}\rangle$
correspond to the excitations of $\hat{\gamma}_{2}$ and  $\hat{\gamma}_{3}$.
Theoretically, the isolated MFs  can be obtained by combinations of these eigenstates
in the limit $\varepsilon _{1,2}\simeq 0$, i.e.,
$|\phi _{\gamma _{1,4}}\rangle =\frac{1}{\sqrt{2}}(|\phi _{\varepsilon
_{1}}\rangle \pm |\phi _{\varepsilon _{-1}}\rangle ),$ and $|\phi _{\gamma
_{2,3}}\rangle =\frac{1}{\sqrt{2}}(|\phi _{\varepsilon _{2}}\rangle \pm
|\phi _{\varepsilon _{-2}}\rangle )$.

\begin{figure}[t]
\includegraphics[width=9.2cm]{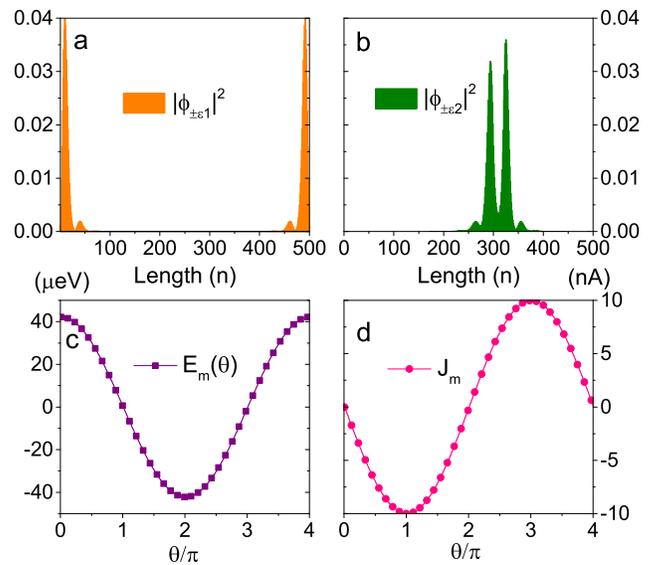}
\caption{\label{fig_2}
(a).  Edge modes $|\phi _{\pm \varepsilon _{1}}|^{2}$ of the lattice model,
correspond to MFs $\hat{\gamma}_{1}$ and $\hat{\gamma}_{4}$.
(b). Edge modes $|\phi _{\pm \varepsilon _{2}}|^{2}$ of the lattice model,
correspond to MFs $\hat{\gamma}_{2}$ and $\hat{\gamma}_{3}$.
(c). Hybridization energy $E_{m}(\theta )$ as a function of the relative field angle $\theta $.
(d). The spin current $J_{m}$ induced by the Magneto-Josephson effects.
 }
\end{figure}

Induced by the overlap of the edge states, the coupling energy between MFs
decays exponentially with the separation \cite{PRB-86-220506}.
According to Ref. \cite{PRL-112-106402},
we now estimate the coupling energy $E_{m}(\theta)$ through the lowest order perturbation theory,
and obtain
\begin{equation}
E_{m}(\theta )\approx \frac{|\langle \phi _{\gamma _{2}}^{0}e^{-i\frac{%
\theta _{l}}{2}\hat{\sigma}_{z}}|H|e^{i\frac{\theta _{r}}{2}\hat{\sigma}%
_{z}}\phi _{\gamma _{3}}^{0}\rangle |}{\sqrt{\langle \phi _{\gamma
_{2}}^{0}|\phi _{\gamma _{2}}^{0}\rangle \langle \phi _{\gamma
_{3}}^{0}|\phi _{\gamma _{3}}^{0}\rangle }}.
\end{equation}
Then, we observe a $4\pi$-periodic dependence
between $E _{m}$ and $\theta $, as shown in Fig.~2(c).
We also calculate the spin current passing through the middle region as
$J_{m}=(e/\hbar )\frac{\partial E^{m}(\theta )}{\partial \theta }$ \cite{NJP-15-015502},
and the result is shown in Fig.~2(d).
The relevant parameters are based on InSb quantum
wires, i.e., $m^{\ast }=0.015$ $m_{e}$, $\alpha=0.2$ eV$\mathring{A}$, $a=10$
nm, $\Delta =0.5$ meV, and $g\mu _{B}=1.5$ meV/T. Other parameters
include $B=200$ mT, $b=200$ mT and $\mu =0$  for the topological regions,
and   $B=0$, $b=200$ mT and $\mu =-0.6$  meV for the non-topological region.
Note that these parameters could  be further optimized.

\subsection{Topology-torsion couplings}

We now discuss the coherent coupling between  the Majorana qubit and the torsional  motion.
Firstly, we combine the MFs to construct two conventional Dirac fermions, and obtain
$\hat{f}_{l}=(\hat{\gamma}_{1}+i\hat{\gamma}_{2})/2$ and $\hat{f}_{r}=(\hat{\gamma}_{3}+i\hat{\gamma}_{4})/2$.
Then, the two quantum states of the Dirac fermions, corresponding to the number of electrons
on either side of the Josephson junction, can function as a physical topological qubit.
These two states are denoted as  $\hat{n}_{x}=\hat{f}_{x}^{\dagger }\hat{f}_{x}=0$ or $1$ $(x=l,r)$,
with parity $\hat{P}_{x}=(-1)^{\hat{f}_{x}^{\dagger }\hat{f}_{x}}=\pm 1$.
As braiding operations can not change the total parity $\hat{P}=\hat{P}_{l}\hat{P}_{r}$ of the system,
four MFs are usually needed to define a logical topological qubit \cite{PRA-87-032327,NJP-16-015009}.
We now define the  topological qubit in the odd parity subspace, and obtain
$|\phi _{TP}\rangle =c_{1}|0\rangle _{l}|1\rangle _{r}+c_{2}|1\rangle
_{l}|0\rangle _{r}$.
% In this odd parity subspace,
For this topological qubit, we also have
\cite{SR-5-12233}
\begin{equation}
i\hat{\gamma}_{1}\hat{\gamma}_{2}\rightarrow -\hat{\sigma}%
_{TP}^{z},i\hat{\gamma}_{2}\hat{\gamma}_{3}\rightarrow -\hat{\sigma}_{TP}^{x},i%
\hat{\gamma}_{3}\hat{\gamma}_{4}\rightarrow \hat{\sigma}_{TP}^{z}.
\end{equation}
%where  $\hat{\sigma}_{TP}^{z}$ and $\hat{\sigma}_{TP}^{x}$  are the spin operators for the  topological qubit.

In this setup, it is the $\theta $ dependence of the hybridization energy $E_{m}(\theta )$ that
leads to the  strong topology-torsion coupling \cite{PRL-112-106402}.
In particular, by expanding $E_{m}(\theta )$ in Eq.~(3) around $\theta_{0}$
to the first order, we arrive at $E_{m}(\theta )=E_{m}(\theta _{0})+\frac{\partial E_{m}(\theta )}{\partial
\theta }\theta _{zpf}(\hat{b}^{\dagger }+\hat{b})$. % with $\theta _{zpf}$ the zero point fluctuations of the torsional mode.
%As the hybridization energy $E_{m}(\theta)$ can be modulated
%by $\theta$ (see Fig.~2 (c)), here
In what follows, we consider the case in which
$E_{m}(\theta _{0})\simeq 0$  (see Fig.~2 (c)), then the Hamiltonian of the topology-torsion dynamics has the form
\begin{equation}
\hat{H}_{Tor-TP}=\hslash \omega _{m}\hat{b}^{\dagger }\hat{b}+\hslash \omega
_{TP}\hat{\sigma}_{TP}^{z}-\hslash g(\hat{b}^{\dagger }+\hat{b})\hat{\sigma}%
_{TP}^{x},
\end{equation}
where $\omega _{TP}=(E_{r}-E_{l})/\hslash$, and $g=\frac{1}{\hslash }\frac{\partial E_{m}(\theta )}
{\partial \theta }\theta_{zpf}$  is the topology-torsion coupling strength.
By tuning the hybridization energy $E_{l}$ and $E_{r}$, we can steer the topology-torsion coupling to
the near-resonance condition $\omega _{TP}\simeq \omega _{m}$.
Under the rotation-wave approximation,
the topology-torsion dynamics can be described by

\begin{equation}
\hat{H}_{Tor-TP}=\hslash \omega _{m}\hat{b}^{\dagger }\hat{b}+\hslash \omega
_{TP}\hat{\sigma}_{TP}^{z}-\hslash g(\hat{b}^{\dagger }\hat{\sigma}_{TP}^{-}+%
\hat{b}\hat{\sigma}_{TP}^{+}).
\end{equation}

\subsection{Spin-torsion couplings}

We now take  into  consideration  the couplings between the mechanical torsional mode and the single NV center.
Before proceeding,  we note that coherent interactions between NV centers and several types of mechanical resonators
have been studied \cite{SCP-56-050303,PRApl-10-024011}. These mechanical resonators include clamped cantilevers \cite{PRB-79-041302}, optically trapped nano-diamond crystals \cite{PRA-96-023827,PRA-96-063810,NC-7-12550}, suspended carbon nanotubes \cite{PRL-112-015502}, etc.
In this setup, the single NV center is positioned straight above the magnetized torsional cantilever, as shown in Fig.~3(a).
When the nanomagnet rotates, the projection of the magnetic field
along the spin component is changed,  resulting in a magnetic coupling between the  torsional mode and the NV center.
The strong magnetic  coupling can be achieved by exquisite preparation of the dressed spin states.

\begin{figure}[t]
\includegraphics[width=7.4cm]{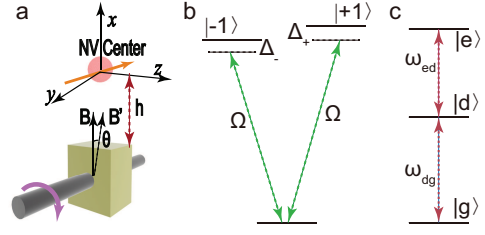}
\caption{\label{fig_3}(Color online) Schematic design and level diagram of
a single NV center coupled to a magnetized torsional cantilever.
(a). The NV center is located  straight above the magnetized torsional cantilever.
The projection of the magnetic field along the
spin component is changed if the torsional cantilever rotates,
leading to the strong spin-torsion coupling.
(b). Simplified energy levels of the NV center in the electronic ground state $|^{3}A_{2}\rangle $.
(c). Dressed spin states in the presence of external driving magnetic fields.}
\end{figure}

An NV center in diamond consists of a substitutional nitrogen atom replacing a carbon atom and an adjacent vacancy.
The electronic ground states of a single NV center are spin triplet states denoted as
$|m_{s}=0,\pm 1\rangle $, and the zero-field splitting  $D_{gs}$ between the degenerate
sublevels $|m_{s}=\pm 1\rangle $  and $|m_{s}=0\rangle $ is $2\pi \times 2.87$ GHz.
For moderate applied magnetic fields, static magnetic field $\vec{B}_{z}=B_{z}\vec{e}_{z}$ causes
Zeeman splitting of the states $|m_{s}=\pm 1\rangle $, while external microwave
field $\vec{B}_{dr}=B_{0}\cos \omega _{0}t\vec{e}_{x}$ polarized in the $x$ direction drives the
states $|m_{s}=0 \rangle $ and  $|m_{s}=\pm 1\rangle $, as shown in Fig.~3(b).
For convenience, we denote as the $z$ axis the crystalline axis of the NV center.

The magnetic field $\vec{B}_{mg}(\theta )$ induced by the suspended nanomagnet depends on the angular displacement $\theta $.
We assume that the NV center is placed in the position  where $\vec{B}_{mg}(\theta )$ is along the $x$
axis, i.e., $\vec{B}_{mg}(\theta )=B_{mg}\cos \theta \cdot \vec{e}%
_{x}+B_{mg}\sin \theta \cdot \vec{e}_{z}$, and $\theta \simeq 0$.
Then, the interaction of the NV center with the total magnetic fields (external driving and from the
nanomagnet) can be described by
\begin{equation}
\hat{H}_{NV}=\hbar D_{gs}\hat{S}_{z}^{2}+g_{e}\mu _{B}[(\vec{B}_{mg}(\theta
)+\vec{B}_{dr})\cdot \hat{\vec{S}}+B_{z}\hat{S}_{z}],
\end{equation}
where $g_{e}=2$ is the NV land\'{e} factor, $\mu _{B}=14\text{MHz}$ $\text{mT}^{-1}$ is
the Bohr magneton, and $\hat{\vec{S}}=(\hat{S}_{x},\hat{S}_{x},\hat{S}_{z})$ is the spin operator of the NV center.
The motion of the  torsional cantilever attached with the nanomagnet produces an angular and time
dependent  magnetic field $\vec{B}_{int}(\theta )\sim \partial _{\theta }\vec{B}_{mg}(\theta )\hat{%
\theta}\cos \omega _{m}t$,
which result in a spin-torsion coupling
interaction $\hat{H}_{int}=g_{e}\mu _{B}\vec{B}_{int}(\theta )\cdot \hat{\vec{S}}\simeq
\hslash \lambda (\hat{b}^{\dagger }+\hat{b})\hat{S}_{z}\cos \omega _{m}t$,
where  the coupling  constant $\lambda =g_{e}\mu _{B}B_{mg}\theta _{zpf}/\hslash $
is proportional to the magnetic field $B_{mg}$
and the zero-point angular extension $\theta _{zpf}$.
In the rotating frame at the frequency $\omega _{m}$, we can acquire the Hamiltonian
to describe the spin-torsion dynamics
\begin{equation}
\hat{H}_{Tor-NV}=\hslash \omega _{m}\hat{b}^{\dagger }\hat{b}+\hat{H}%
_{NV}+\hslash \lambda (\hat{b}^{\dagger }+\hat{b})\hat{S}_{z}.
\end{equation}

Note that the far-off resonant interactions between the NV spin and the static and  low frequency components of magnetic fields
along the  $x$  axis can be ignored. In the basis defined by the eigenstates of $\hat{S}_{z}$, i.e., $\{|m_{s}\rangle ,m_{s}=0,\pm 1\},$
with $\hat{S}_{z}|m_{s}\rangle =m_{s}|m_{s} \rangle$, we have
\begin{eqnarray}
\hat{H}_{NV} &=&\sum\limits_{m_{s}}[\hbar D_{gs}m_{s}^{2}+g_{e}\mu
_{B}(B_{z}+B_{mg}\sin \theta )m_{s}]|m_{s}\rangle \langle m_{s}|  \nonumber
\\
&&+\sum\limits_{m_{s}m_{s^{\prime }}}g_{e}\mu _{B}B_{0}\cos \omega
_{0}t\langle m_{s}|\hat{S}_{x}|m_{s}^{^{\prime }}\rangle |m_{s}\rangle
\langle m_{s}^{^{\prime }}|,
\end{eqnarray}
where $\hat{S}_{x}=\frac{\hslash }{\sqrt{2}}(|0\rangle \langle +1|+|0\rangle\langle -1|+H.c.)$.
In the rotating-frame at the driving frequency  $\omega _{0}$ and under the rotating-wave approximation,
we get
\begin{eqnarray}
\hat{H}_{NV} &=&\hslash \Delta _{+}|+1\rangle \langle +1|+\hslash \Delta
_{-}|+1\rangle \langle +1|+  \nonumber \\
&&\hslash \Omega (|-1\rangle \langle 0|+|+1\rangle \langle 0|+H.c.)
\end{eqnarray}
where $\hslash \Delta _{\pm }=\hslash D_{gs}\pm g_{e}\mu _{B}(B_{z}\pm B_{mg}\sin
\theta )-\hslash \omega _{0}$, and $\hslash \Omega =\frac{\sqrt{2}}{4}g_{e}\mu
_{B}B_{0}$.

In the following we consider the symmetric detuning condition $\Delta _{+}=\Delta _{-}=\Delta $ for simplicity
(e.g., $B_{z}\rightarrow 0$).
We now define the bright spin state   $|b\rangle =(|+1\rangle +|-1\rangle )/\sqrt{2}$
 and the dark spin state  $|d\rangle =(|+1\rangle -|-1\rangle )/\sqrt{2}$.
Then we find that the Hamiltonian (12) couples
the state $|0\rangle $ to the bright state $|b\rangle $, while the dark state $|d\rangle $ remains decoupled.
The resulting eigenstates of $\hat{H}_{NV}$ are therefore given by $|d\rangle$ and two dressed states
$|g\rangle =\cos (\alpha )|0\rangle -\sin (\alpha )|b\rangle $ and $|e\rangle =\sin (\alpha )|0\rangle +\cos (\alpha )|b\rangle $,
with  $\tan (2\alpha )=2\sqrt{2}\Omega /\Delta $. The corresponding eigenfrequencies  are given by $\omega _{d}=\Delta $, $\omega _{e/g}=(\Delta \pm \sqrt{\Delta ^{2}+8\Omega
^{2}})/2$,
as illustrated in Fig.~3(c).

In the dressed state basis $\{|g\rangle ,|d\rangle ,|e\rangle \}$, the
Hamiltonian (10) can be rewritten as
\begin{eqnarray}
\hat{H}_{Tor-NV} &=&\hslash \omega _{m}\hat{b}^{\dagger }\hat{b}+\hslash
\omega _{eg}|e\rangle \langle e|+\hslash \omega _{dg}|d\rangle \langle d|+\nonumber \\
&&\hslash (\lambda _{g}|g\rangle \langle d|+\lambda _{e}|d\rangle \langle
e|+H.c.)(\hat{b}^{\dagger }+\hat{b}),\nonumber \\
\end{eqnarray}
where $\lambda _{g}=-\lambda \sin (\alpha )$ and $\lambda _{e}=\lambda \cos(\alpha )$.
We further adjust the values of $B_{0}$ to drive the system to the
resonance condition $\omega _{ed}\simeq \omega _{m}\ll \omega _{dg}$, and
then the far-off-resonance state $|g\rangle $ can be neglected.
Under the rotating-wave approximation,
the spin-torsion dynamics can be described by the JC interaction Hamiltonian
\begin{equation}
\hat{H}_{Tor-NV}=\hslash \omega _{m}\hat{b}^{\dagger }\hat{b}+\hslash \omega
_{NV}\hat{\sigma}_{NV}^{z}+\hslash \lambda _{e}(\hat{b}^{\dagger }\hat{\sigma%
}_{NV}^{-}+\hat{b}\hat{\sigma}_{NV}^{+}),
\end{equation}
where
$\hat{\sigma}_{NV}^{-}=|d\rangle \langle e|,\hat{\sigma}_{NV}^{+}=|e\rangle
\langle d|$,  $\hat{\sigma}_{NV}^{z}=|e\rangle \langle e|-|d\rangle \langle d|$,
and $\omega _{NV}=\omega _{e}-\omega _{d}$.

\section{EXPERIMENTAL PARAMETERS}

According to the Hamiltonian (8) and (14), the whole system can be described  by
\begin{eqnarray}
\hat{H}_{sys} &=&\hslash \omega _{TP}\hat{\sigma}_{TP}^{z}+\hslash \omega
_{m}\hat{b}^{\dagger }\hat{b}+\hslash \omega _{NV}\hat{\sigma}_{NV}^{z}
\nonumber \\
&&-\hslash g(t)\hat{b}^{\dagger }\hat{\sigma}_{TP}^{-}+\hslash \lambda
_{e}(t)\hat{b}^{\dagger }\hat{\sigma}_{NV}^{-}+H.c..
\end{eqnarray}
The first three terms describe the free Hamiltonian of the system,
while the last four terms describe two JC interactions:
one between the  torsional mode and the topological qubit, and the other between the torsional mode and the single NV center.
As the torsional cantilever couples to the Majorana qubit and the NV center simultaneously,
it is possible to  work as a quantum interface for quantum state conversion.

Let us discuss the dynamics of the system in a realistic situation.
For this setup,  the relaxation ($\Gamma _{1}$) and dephasing ($\Gamma _{2}$) of the Majorana qubit,
the decay of the mechanical resonator ($\gamma _{m}$), and also the dephasing of the NV center ($\gamma _{s}$)
should be taken into consideration. As a result, the full dynamics of this system can be described by the following master equation
\begin{eqnarray}
\frac{d\hat{\rho}(t)}{dt} &=&-\frac{i}{\hslash  }[\hat{H}_{sys},\hat{\rho}%
]+\gamma _{s}\mathcal{D}(\hat{\sigma}_{NV}^{z})  \nonumber \\
&&+\Gamma _{1}\mathcal{D(}\hat{\sigma}_{TP}^{-})+\Gamma _{2}\mathcal{D}(\hat{%
\sigma}_{TP}^{+}\hat{\sigma}_{TP}^{-}) \\
&&+(n_{th}+1)\gamma _{m}\mathcal{D}(\hat{b})+n_{th}\gamma _{m}\mathcal{D}(%
\hat{b}^{\dagger }),  \nonumber
\end{eqnarray}
where $n_{th}=(e^{\hslash  \omega _{m}/k_{B}T}-1)^{-1}$ is
the thermal phonon number at the environment temperature $T$,
and $\mathcal{D}(\hat{o})=\hat{o}\hat{\rho}\hat{o}^{\dagger }-\frac{1}{2}\hat{o}%
^{\dagger }\hat{o}\hat{\rho}-\frac{1}{2}\hat{\rho}\hat{o}^{\dagger }\hat{o}$
for a given operator $\hat{o}$.

\begin{figure}[b]
\includegraphics[width=7.4cm]{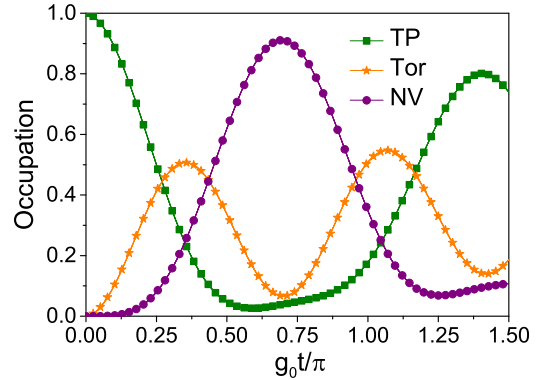}
\caption{\label{fig_4}(Color online) Rabi oscillations of the hybrid tripartite system
interfacing a topological qubit (TP) and a single NV center  (NV) via
 a magnetized torsional cantilever (Tor).
The topological qubit is initially in state $|1\rangle$,
while the torsional mode is initially in the ground state
and the NV center  in the state $|d\rangle$.
We consider the system in a realistic case, with the
coupling parameters $g=\lambda _{g}=g_{0}\sim 2\pi\times 200$ kHz, and the decay parameters
$\Gamma _{1}=\Gamma _{2}=0.05g_{0}$, $\gamma _{m}=0.0002g_{0}$, $n_{th}=104$,
and $\gamma _{s}=0.1g_{0}$.
}
\end{figure}

We now consider the  experimental feasibility of our scheme and the appropriate parameters to achieve strong couplings.
This manuscript employs the magnetized torsional cantilever to interface the
Majorana qubit and the conventional spin qubit. As an  alternative,
one may also use the edge states of topological insulators
or the planar Josephson junctions \cite{arxiv-1089-03037v1} to realize the Majorana qubit, which requires weaker magnetic fields.
For the magnetized torsional cantilever, the nanomagnet suspended on the single-walled nanotube has a
size of $80\times 40\times 20$ nm$^{3}$, and a moment of inertia of $I\sim 4.8\times 10^{-33}$ kgm$^{2}$
with respect to the tube axis \cite{SC-309-1539}.
Then the  torsional cantilever has a resonance frequency  $\omega _{m}=[\frac{1}{2\pi }][(\frac{K}{I})^{\frac{1}{2}}]\sim 2\pi\times4$ MHz,
and an angle of zero point fluctuations $\theta _{zpf}=(\hslash  ^{2}/KI)^{1/4}\sim 3\times 10^{-5}$, with
$K \sim 3\times 10^{-18}$ Nm per radin  the torsional spring constant of the tube axis.
As shown in Fig.~2(c), the topology-torsion coupling energy is about $E_{m}(\theta \simeq 0)\sim 42$
$\mu$ eV, which implies a coupling constant $g\sim 2\pi\times 200$ kHz.
Moreover, the spin-torsion coupling can be  modified by the distance between the torsional
cantilever and the NV center. For a distance $h\sim 80$ nm, one can obtain $B_{mg}\sim 80$ mT \cite{NatNano-2-301} and
$\lambda _{e}\sim 2\pi\times 200$ kHz.

We now take the decoherence processes of the system into consideration. In practical situations,
the relaxation and dephasing rates of Majorana qubit are strongly depended on the concrete realization,
and here we take $\Gamma _{1}\simeq \Gamma _{2}\sim 2\pi\times 10$ kHz \cite{NP-7-412,PRB-85-020502}.
When it comes to the mechanical damping,
the recent fabrication of carbon nanotube resonators can possess quality factors exceeding $10^{5}$.
For a torsional cantilever with $\omega _{m}\sim 2\pi\times 4$ MHz,
the mechanical damping rate is about $\gamma _{m}=\omega _{m}/Q\sim 2\pi\times 40$ Hz, and
the the thermal phonon occupation number is about  $n_{th}\sim 104$ at the temperature of $T \sim 20$ mK.
As for the NV center, the dephasing time $T_{2}$ as long as several milliseconds can be reached in ultrapure
diamond \cite{prb-82-201201},  and here we take $\gamma _{s}\sim 2\pi\times 2$ kHz.
Finally, with these realistic parameters, we perform numerical simulations for the quantum dynamics of the system
by solving the master equation (16). As shown in Fig.~4, the coherent interactions can dominate
the decoherence processes in this hybrid quantum device, which enables the system to enter the strong coupling regime.

\section{Dark state conversion}

In the previous sections, it is  shown that the magnetized torsional cantilever can be employed as an
intermediary  to link the topological qubit and the single NV center.
Utilizing this quantum interface, one of the applications is to transfer the quantum state
from the topological qubit to the spin qubit. The efficiency of the transformation relies
on the specific protocol for quantum state conversion.
To transfer quantum state in this setup,
we show that both the direct-transfer scheme and the dark-state protocol are available.

As shown in Eq. (15), this hybrid tripartite system can be modeled by a beam-splitter Hamiltonian
composed of two JC interactions. Therefore, we can use the
JC interactions for quantum state conversion. In particular, this direct-transfer  scheme usually includes two steps:
step 1, turn on the topology-torsion couplings  $g$ for a time $t_{1}=\frac{\pi }{2g}$
(while $\lambda _{e}=0$), to transfer the quantum state from the Majorana qubit to the torsional mode; step 2,
turn off $g$ and turn on the spin-torsion couplings $\lambda _{e}$  for a time $t_{2}=\frac{\pi }{2\lambda _{e}}$
(while $g =0$), to map the quantum state from the torsional mode to the spin qubit.
Note that this direct-transfer scheme has been studied in optomechanical system \cite{pra-82-053806}.
In our setup, the topology-torsion couplings $g$ can be controlled by
the electrostatic gates \cite{NP-7-412},
while the spin-torsion couplings $\lambda _{e}$ can be controlled by the  external driving fields.

As the direct-transfer scheme may suffer from decoherence of the mechanical mode,
here we propose to realize the coherent quantum state conversion via a dark-state protocol
\cite{PRL-108-153603, PRL-108-153604,prapl-04-044003}.
This dark-state protocol is particularly robust against mechanical
dissipations, as it decoupled from the vibrational mode.
%In the following we assume that the system is under the resonance condition
We proceed by assuming that
$\omega _{m}\simeq \omega _{TP}\simeq \omega _{NV}$.
To describe quasiparticles formed by combinations of spin and Majorana excitations,
we now introduce two polariton operators $\hat{c}_{br}=\sin (\beta )\hat{\sigma}_{TP}^{-}+\cos (\beta )\hat{\sigma}%
_{NV}^{-}$, and $\hat{c}_{dk}=-\cos (\beta )\hat{\sigma}_{TP}^{-}+\sin
(\beta )\hat{\sigma}_{NV}^{-}$, with $tan(\beta )=-g/\lambda _{e}$.
Then, one can verify that the Hamiltonian (15) can take a compact form
\begin{equation}
\hat{H}_{sys}=\hslash \tilde{\omega}_{+}\hat{c}_{+}^{\dagger }\hat{c}%
_{+}+\hslash \tilde{\omega}_{-}\hat{c}_{-}^{\dagger }\hat{c}_{-}+\hslash
\tilde{\omega}_{dk}\hat{c}_{dk}^{\dagger }\hat{c}_{dk},
\end{equation}
where $\hat{c}_{\pm }=(1/\sqrt{2})(\hat{c}_{br}\pm \hat{b})$
%being the Majorana-phonon-spin polaron operators.
describe polarons formed by combinations of polariton and phonon excitations.
The corresponding frequencies of the polarons
are $\tilde{\omega}_{d}=\omega _{m}$, $\tilde{\omega}_{\pm }=\omega _{m}\pm
\sqrt{\lambda _{e}^{2}+g^{2}}$.
We refer to $\hat{c}_{dk}$ as the mechanically dark polariton operator,
as it involves the spin and Majorana operations only, decoupled from the
 torsional mode.

\begin{figure}[t]
\includegraphics[width=8.0cm]{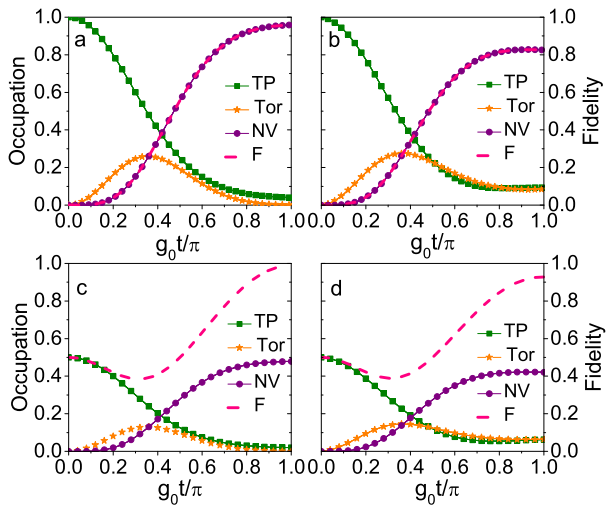}
\caption{\label{fig_dark_1}(Color online) Fidelity (F) and occupations (TP,Tor,NV) as
a function of time in the dark-state protocol,
with the coupling parameters $g(t)=g_{0}e^{-(t-\pi )^{2}/30}$,
and $\lambda _{e}(t)=1.5g_{0}e^{-t^{2}/6}$.
Two kinds of initial states are under consideration:
(\romannumeral1). a Fock state $|1\rangle$ in (a) and (b); (\romannumeral2). a superposition state
$\frac{1}{\sqrt{2}}|0\rangle +\frac{1}{\sqrt{2}}|1\rangle $ in (c) and (d).
The decoherence parameters for (a) and (c)  are chosen as
$\Gamma _{1}=\Gamma _{2}=\gamma _{m}=\gamma _{s}=0$,
and for (b) and (d)
$\Gamma _{1}=\Gamma _{2}=0.05g_{0}$, $\gamma _{m}=0.0002g_{0}$, $n_{th}=104$,
and $\gamma _{s}=0.1g_{0}$.}
\end{figure}

The excitation of $\hat{c}_{dk}$ means  the system remains in the mechanically dark state.
Utilizing the adiabatic evolution of the dark state, we can transfer the quantum state
from the  topological qubit to the single NV center.
In particular, in the limit  $\beta = 0$, we get
$\hat{c}_{dk}=-\hat{\sigma}_{TP}^{-}$, while in the limit $\beta = \pi /2$, we get $\hat{c}_{dk}=\hat{\sigma}_{NV}^{-}$.
This implies, if we adiabatically rotate the mixing angle
$\beta $ form zero to $\pi /2$, the mechanically dark polariton $\hat{c}_{dk}$ would
evolve from $-\hat{\sigma}_{TP}^{-}$ to $\hat{\sigma}_{NV}^{-}$.
At the same time, the quantum state of  the topological qubit could be transferred to the
spin qubit except for a phase factor $e^{-i\pi }$.

Similar to the well-known stimulated Raman adiabatic passage,
the dark state protocol can be implemented by an adiabatic passage approach.
To transfer the quantum state from the topological qubit to the NV center,
we initially make $\lambda _{e}(t)$ larger than $g(t)$, but keep $g(t)$ finite.
Then, we modulate the coupling parameters carefully so that the
dark polariton operator evolves adiabatically from
 $-\hat{\sigma}_{TP}^{-}$ at the beginning to $\hat{\sigma}_{NV}^{-}$  at the end.
Meanwhile, the quantum state of the topological qubit will be
transferred to the NV center as the system evolves.
To satisfy the adiabatic conditions, one need to modulate the coupling  strengths slowly
to ensure that the system adiabatically follows the dark polaritons.

We perform numerical simulations for the dark-state protocol by solving the
master equation (16) with Hamiltonian (15),
where the coupling parameters $g(t)=g_{0}e^{-(t-\pi )^{2}/30}$,
$\lambda _{e}(t)=1.5g_{0}e^{-t^{2}/6}$,
and $g_{0}\sim 2\pi \times 200$ kHz as discussed in Sec. \uppercase\expandafter{\romannumeral3}.
To test the robustness of the protocol,
we consider two kinds of initial states for the topological qubit:
a Fock state $|1\rangle $ as displayed in Figs.~5(a) and 5(b), and a superposition state
$\frac{1}{\sqrt{2}}|0\rangle+\frac{1}{\sqrt{2}}|1\rangle$
as displayed in Figs.~5(c) and 5(d).
For the ideal case without decoherence, high fidelities 0.96, 0.99 can be reached,
as shown in Figs.~5(a) and 5(c).
We find that,
as the system evolves, the quantum states of the topological qubit
is slowly transferred to the NV center spin in the conversion process.
Then, when it turns to the realistic case,
high fidelities 0.83 and 0.93 can be reached, as shown in Figs.~5(b) and 5(d).
Therefore, this dark-state protocol works very well in practical conditions.

\section{CONCLUSIONS}

In summary, we present a hybrid quantum device interfacing a
topological qubit and a single NV center via a high-$Q$ magnetized torsional cantilever.
We show that, the mechanical torsional mode can strongly couple to
the Majorana qubit and the single NV center simultaneously, and then interface them for quantum states conversion.
In particular, the topology-torsion couplings are induced by the spin currents passing through the TNT junction,
while the spin-torsion couplings can be realized  by the exquisite preparation of dressed spin states and using a suspended nanomagnet.
We also find that, under resonance conditions, one eigenstate
of the system is the mechanically dark state, which provides a dark-state protocol for quantum states conversion
with very high fidelities. As the mechanical mode keeps unpopulated during the conversion process,
this dark-state scheme is extremely robust against mechanical damping.
This hybrid quantum architecture integrated with the topological
qubits and NV spins provides a potential platform for quantum information processing.

\section*{Acknowledgments}

B.L. thanks  Di-Hao Sun and Shuai Li for fruitful discussions.
This work was support by  the NSFC under Grants  No. 11774285, 91536115, 11534008, and the Fundamental Research Funds for the Central Universities. Part of the simulations are coded in PYTHON using the QUTIP library \cite{CPC}.

%\bibliography{TPRef}

\begin{thebibliography}{84}%
\makeatletter
\providecommand \@ifxundefined [1]{%
 \@ifx{#1\undefined}
}%
\providecommand \@ifnum [1]{%
 \ifnum #1\expandafter \@firstoftwo
 \else \expandafter \@secondoftwo
 \fi
}%
\providecommand \@ifx [1]{%
 \ifx #1\expandafter \@firstoftwo
 \else \expandafter \@secondoftwo
 \fi
}%
\providecommand \natexlab [1]{#1}%
\providecommand \enquote  [1]{``#1''}%
\providecommand \bibnamefont  [1]{#1}%
\providecommand \bibfnamefont [1]{#1}%
\providecommand \citenamefont [1]{#1}%
\providecommand \href@noop [0]{\@secondoftwo}%
\providecommand \href [0]{\begingroup \@sanitize@url \@href}%
\providecommand \@href[1]{\@@startlink{#1}\@@href}%
\providecommand \@@href[1]{\endgroup#1\@@endlink}%
\providecommand \@sanitize@url [0]{\catcode `\\12\catcode `\$12\catcode
  `\&12\catcode `\#12\catcode `\^12\catcode `\_12\catcode `\%12\relax}%
\providecommand \@@startlink[1]{}%
\providecommand \@@endlink[0]{}%
\providecommand \url  [0]{\begingroup\@sanitize@url \@url }%
\providecommand \@url [1]{\endgroup\@href {#1}{\urlprefix }}%
\providecommand \urlprefix  [0]{URL }%
\providecommand \Eprint [0]{\href }%
\providecommand \doibase [0]{https://doi.org/}%
\providecommand \selectlanguage [0]{\@gobble}%
\providecommand \bibinfo  [0]{\@secondoftwo}%
\providecommand \bibfield  [0]{\@secondoftwo}%
\providecommand \translation [1]{[#1]}%
\providecommand \BibitemOpen [0]{}%
\providecommand \bibitemStop [0]{}%
\providecommand \bibitemNoStop [0]{.\EOS\space}%
\providecommand \EOS [0]{\spacefactor3000\relax}%
\providecommand \BibitemShut  [1]{\csname bibitem#1\endcsname}%
\let\auto@bib@innerbib\@empty
%</preamble>
\bibitem [{\citenamefont {Elliott}\ and\ \citenamefont
  {Franz}(2015)}]{RMP-87-137}%
  \BibitemOpen
  \bibfield  {author} {\bibinfo {author} {\bibfnamefont {S.~R.}\
  \bibnamefont {Elliott}}\ and\ \bibinfo {author} {\bibfnamefont {M.}\
  \bibnamefont {Franz}},\ }\bibfield  {title} {\bibinfo {title} {Colloquium:
  Majorana fermions in nuclear, particle, and solid-state physics},\ }\href
  {https://doi.org/10.1103/RevModPhys.87.137} {\bibfield  {journal} {\bibinfo
  {journal} {Rev. Mod. Phys.}\ }\textbf {\bibinfo {volume} {87}},\ \bibinfo
  {pages} {137} (\bibinfo {year} {2015})}\BibitemShut {NoStop}%
\bibitem [{\citenamefont {Lutchyn}\ \emph {et~al.}(2018)\citenamefont
  {Lutchyn}, \citenamefont {Bakkers}, \citenamefont {Kouwenhoven},
  \citenamefont {Krogstrup}, \citenamefont {Marcus},\ and\ \citenamefont
  {Oreg}}]{NR-3-52}%
  \BibitemOpen
  \bibfield  {author} {\bibinfo {author} {\bibfnamefont {R.~M.}\ \bibnamefont
  {Lutchyn}}, \bibinfo {author} {\bibfnamefont {E.~P. A.~M.}\ \bibnamefont
  {Bakkers}}, \bibinfo {author} {\bibfnamefont {L.~P.}\ \bibnamefont
  {Kouwenhoven}}, \bibinfo {author} {\bibfnamefont {P.}~\bibnamefont
  {Krogstrup}}, \bibinfo {author} {\bibfnamefont {C.~M.}\ \bibnamefont
  {Marcus}},\ and\ \bibinfo {author} {\bibfnamefont {Y.}~\bibnamefont {Oreg}},\
  }\bibfield  {title} {\bibinfo {title} {Majorana zero modes in
  superconductor csemiconductor heterostructures},\ }\href
  {https://www.nature.com/articles/s41578-018-0003-1} {\bibfield  {journal}
  {\bibinfo  {journal} {Nat. Rev. Mater.}\ }\textbf {\bibinfo {volume} {3}},\
  \bibinfo {pages} {52} (\bibinfo {year} {2018})}\BibitemShut {NoStop}%
\bibitem [{\citenamefont {Hao}\ \emph {et~al.}(2018)\citenamefont {Hao},
  \citenamefont {Chun-Xiao}, \citenamefont {Sasa}, \citenamefont {Di},
  \citenamefont {Logan}, \citenamefont {Wang}, \citenamefont {van Loo},
  \citenamefont {Bommer}, \citenamefont {de~Moor}, \citenamefont {Diana},
  \citenamefont {het Veld}, \citenamefont {van Veldhoven}, \citenamefont
  {Sebastian}, \citenamefont {Verheijen}, \citenamefont {Mihir}, \citenamefont
  {Pennachio}, \citenamefont {Shojaei}, \citenamefont {Lee}, \citenamefont
  {Palmstrom}, \citenamefont {Bakkers}, \citenamefont {Sarma},\ and\
  \citenamefont {Kouwenhoven}}]{Natrue-74-556}%
  \BibitemOpen
  \bibfield  {author} {\bibinfo {author} {\bibfnamefont {H.}\ \bibnamefont
  {Zhang}}, \bibinfo {author} {\bibfnamefont {C.-X.}\ \bibnamefont {Liu}},
  \bibinfo {author} {\bibfnamefont {S.}\ \bibnamefont {Gazibegovic}},
  \bibinfo {author} {\bibfnamefont {D.}~\bibnamefont {Xu}}, \bibinfo {author}
  {\bibfnamefont {J.~A.}\ \bibnamefont {Logan}}, \bibinfo {author}
  {\bibfnamefont {G.-Z.}\ \bibnamefont {Wang}}, \bibinfo {author}
  {\bibfnamefont {N.}\ \bibnamefont {van Loo}}, \bibinfo {author}
  {\bibfnamefont {J.~D.~S.}\ \bibnamefont {Bommer}}, \bibinfo {author}
  {\bibfnamefont {M.~W.~A.}\ \bibnamefont {de~Moor}}, \bibinfo {author}
  {\bibfnamefont {D.}\ \bibnamefont {Car}}, \bibinfo {author} {\bibfnamefont
  {R.~L.~M.~Op}\ \bibnamefont {het Veld}}, \bibinfo {author} {\bibfnamefont
  {P.~J.}\ \bibnamefont {van Veldhoven}}, \bibinfo {author} {\bibfnamefont
  {S.}\ \bibnamefont {Koelling}}, \bibinfo {author} {\bibfnamefont
  {M.~A.}\ \bibnamefont {Verheijen}}, \bibinfo {author} {\bibfnamefont
  {M.}\ \bibnamefont {Pendharkar}}, \bibinfo {author} {\bibfnamefont
  {D.~J.}\ \bibnamefont {Pennachio}}, \bibinfo {author} {\bibfnamefont
  {B.}\ \bibnamefont {Shojaei}}, \bibinfo {author} {\bibfnamefont
  {J.~Sue}\ \bibnamefont {Lee}}, \bibinfo {author} {\bibfnamefont {C.~J.}\
  \bibnamefont {Palmstrom}}, \bibinfo {author} {\bibfnamefont {E.~P.~A.~M.}\
  \bibnamefont {Bakkers}}, \bibinfo {author} {\bibfnamefont {S.~Das}\
  \bibnamefont {Sarma}}, \ and\ \bibinfo {author} {\bibfnamefont {L.~P.}\
  \bibnamefont {Kouwenhoven}},\ }\bibfield  {title}  {\bibinfo {title}
  {Quantized Majorana conductance},\ }\href {\doibase 10.1038/nature26142}
  {\bibfield  {journal} {\bibinfo  {journal} {Nature}\ }\textbf
  {\bibinfo {volume} {556}},\ \bibinfo {pages} {74} (\bibinfo {year}
  {2018})}\BibitemShut {NoStop}%
\bibitem [{\citenamefont {Wilczek}(2009)}]{NP-5-614}%
  \BibitemOpen
  \bibfield  {author} {\bibinfo {author} {\bibfnamefont {F.}~\bibnamefont
  {Wilczek}},\ }\bibfield  {title} {\bibinfo {title} {Majorana returns},\
  }\href {https://www.nature.com/articles/nphys1380} {\bibfield  {journal}
  {\bibinfo  {journal} {Nature Phys.}\ }\textbf {\bibinfo {volume} {5}},\
  \bibinfo {pages} {614} (\bibinfo {year} {2009})}\BibitemShut {NoStop}%
\bibitem [{\citenamefont {Litinski}\ and\ \citenamefont {von
  Oppen}(2018)}]{PRB-97-205404}%
  \BibitemOpen
  \bibfield  {author} {\bibinfo {author} {\bibfnamefont {D.}\ \bibnamefont
  {Litinski}}\ and\ \bibinfo {author} {\bibfnamefont {F.}\ \bibnamefont {von
  Oppen}},\ }\bibfield  {title} {\bibinfo {title} {Quantum computing with
  Majorana fermion codes},\ }\href {https://doi.org/10.1103/PhysRevB.97.205404}
  {\bibfield  {journal} {\bibinfo  {journal} {Phys. Rev. B}\ }\textbf {\bibinfo
  {volume} {97}},\ \bibinfo {pages} {205404} (\bibinfo {year}
  {2018})}\BibitemShut {NoStop}%
\bibitem [{\citenamefont {Read}\ and\ \citenamefont
  {Green}(2000)}]{PRB-61-10267}%
  \BibitemOpen
  \bibfield  {author} {\bibinfo {author} {\bibfnamefont {N.}~\bibnamefont
  {Read}}\ and\ \bibinfo {author} {\bibfnamefont {D.}\ \bibnamefont
  {Green}},\ }\bibfield  {title} {\bibinfo {title} {Paired states of fermions
  in two dimensions with breaking of parity and time-reversal symmetries and
  the fractional quantum Hall effect},\ }\href
  {https://doi.org/10.1103/PhysRevB.61.10267} {\bibfield  {journal} {\bibinfo
  {journal} {Phys. Rev. B}\ }\textbf {\bibinfo {volume} {61}},\ \bibinfo
  {pages} {10267} (\bibinfo {year} {2000})}\BibitemShut {NoStop}%
\bibitem [{\citenamefont {Fu}\ and\ \citenamefont
  {Kane}(2008)}]{PRL-100-096407}%
  \BibitemOpen
  \bibfield  {author} {\bibinfo {author} {\bibfnamefont {L.}\ \bibnamefont
  {Fu}}\ and\ \bibinfo {author} {\bibfnamefont {C.~L.}\ \bibnamefont {Kane}},\
  }\bibfield  {title} {\bibinfo {title} {Superconducting proximity effect and
  Majorana fermions at the surface of a topological insulator},\ }\href
  {https://doi.org/10.1103/PhysRevLett.100.096407} {\bibfield  {journal}
  {\bibinfo  {journal} {Phys. Rev. Lett.}\ }\textbf {\bibinfo {volume} {100}},\
  \bibinfo {pages} {096407} (\bibinfo {year} {2008})}\BibitemShut {NoStop}%
\bibitem [{\citenamefont {Nayak}\ \emph {et~al.}(2008)\citenamefont {Nayak},
  \citenamefont {Simon}, \citenamefont {Stern}, \citenamefont {Freedman},\ and\
  \citenamefont {Das~Sarma}}]{RevModPhys-80-1083}%
  \BibitemOpen
  \bibfield  {author} {\bibinfo {author} {\bibfnamefont {C.}\ \bibnamefont
  {Nayak}}, \bibinfo {author} {\bibfnamefont {S.~H.}\ \bibnamefont
  {Simon}}, \bibinfo {author} {\bibfnamefont {A.}\ \bibnamefont {Stern}},
  \bibinfo {author} {\bibfnamefont {M.}\ \bibnamefont {Freedman}},\ and\
  \bibinfo {author} {\bibfnamefont {S.}\ \bibnamefont {Das~Sarma}},\
  }\bibfield  {title} {\bibinfo {title} {Non-abelian anyons and topological
  quantum computation},\ }\href {https://doi.org/10.1103/RevModPhys.80.1083}
  {\bibfield  {journal} {\bibinfo  {journal} {Rev. Mod. Phys.}\ }\textbf
  {\bibinfo {volume} {80}},\ \bibinfo {pages} {1083} (\bibinfo {year}
  {2008})}\BibitemShut {NoStop}%
\bibitem [{\citenamefont {Sau}\ \emph {et~al.}(2010)\citenamefont {Sau},
  \citenamefont {Lutchyn}, \citenamefont {Tewari},\ and\ \citenamefont
  {Das~Sarma}}]{PRL-104-040502}%
  \BibitemOpen
  \bibfield  {author} {\bibinfo {author} {\bibfnamefont {J.~D.}\ \bibnamefont
  {Sau}}, \bibinfo {author} {\bibfnamefont {R.~M.}\ \bibnamefont {Lutchyn}},
  \bibinfo {author} {\bibfnamefont {S.}\ \bibnamefont {Tewari}},\ and\
  \bibinfo {author} {\bibfnamefont {S.}~\bibnamefont {Das~Sarma}},\ }\bibfield
  {title} {\bibinfo {title} {Generic new platform for topological quantum
  computation using semiconductor heterostructures},\ }\href
  {https://doi.org/10.1103/PhysRevLett.104.040502} {\bibfield  {journal}
  {\bibinfo  {journal} {Phys. Rev. Lett.}\ }\textbf {\bibinfo {volume} {104}},\
  \bibinfo {pages} {040502} (\bibinfo {year} {2010})}\BibitemShut {NoStop}%
\bibitem [{\citenamefont {Alicea}(2010)}]{PRB-81-125318}%
  \BibitemOpen
  \bibfield  {author} {\bibinfo {author} {\bibfnamefont {J.}\ \bibnamefont
  {Alicea}},\ }\bibfield  {title} {\bibinfo {title} {Majorana fermions in a
  tunable semiconductor device},\ }\href
  {https://doi.org/10.1103/PhysRevB.81.125318} {\bibfield  {journal} {\bibinfo
  {journal} {Phys. Rev. B}\ }\textbf {\bibinfo {volume} {81}},\ \bibinfo
  {pages} {125318} (\bibinfo {year} {2010})}\BibitemShut {NoStop}%
\bibitem [{\citenamefont {Lutchyn}\ \emph {et~al.}(2010)\citenamefont
  {Lutchyn}, \citenamefont {Sau},\ and\ \citenamefont
  {Das~Sarma}}]{PRL-105-077001}%
  \BibitemOpen
  \bibfield  {author} {\bibinfo {author} {\bibfnamefont {R.~M.}\
  \bibnamefont {Lutchyn}}, \bibinfo {author} {\bibfnamefont {J.~D.}\
  \bibnamefont {Sau}},\ and\ \bibinfo {author} {\bibfnamefont {S.}~\bibnamefont
  {Das~Sarma}},\ }\bibfield  {title} {\bibinfo {title} {Majorana fermions and a
  topological phase transition in semiconductor-superconductor
  heterostructures},\ }\href {https://doi.org/10.1103/PhysRevLett.105.077001}
  {\bibfield  {journal} {\bibinfo  {journal} {Phys. Rev. Lett.}\ }\textbf
  {\bibinfo {volume} {105}},\ \bibinfo {pages} {077001} (\bibinfo {year}
  {2010})}\BibitemShut {NoStop}%
\bibitem [{\citenamefont {Chen}\ \emph {et~al.}(2014)\citenamefont {Chen},
  \citenamefont {LiMing}, \citenamefont {Huang}, \citenamefont {Yin},\ and\
  \citenamefont {Xue}}]{PRA-90-012323}%
  \BibitemOpen
  \bibfield  {author} {\bibinfo {author} {\bibfnamefont {L.}\ \bibnamefont
  {Chen}}, \bibinfo {author} {\bibfnamefont {W.}~\bibnamefont {LiMing}},
  \bibinfo {author} {\bibfnamefont {J.-H.}\ \bibnamefont {Huang}}, \bibinfo
  {author} {\bibfnamefont {C.-P.}\ \bibnamefont {Yin}},\ and\ \bibinfo
  {author} {\bibfnamefont {Z.-Y.}\ \bibnamefont {Xue}},\ }\bibfield
  {title} {\bibinfo {title} {Majorana zero modes on a one-dimensional chain for
  quantum computation},\ }\href {https://doi.org/10.1103/PhysRevA.90.012323}
  {\bibfield  {journal} {\bibinfo  {journal} {Phys. Rev. A}\ }\textbf {\bibinfo
  {volume} {90}},\ \bibinfo {pages} {012323} (\bibinfo {year}
  {2014})}\BibitemShut {NoStop}%
\bibitem [{\citenamefont {Oreg}\ \emph {et~al.}(2010)\citenamefont {Oreg},
  \citenamefont {Refael},\ and\ \citenamefont {von Oppen}}]{PRL-105-177002}%
  \BibitemOpen
  \bibfield  {author} {\bibinfo {author} {\bibfnamefont {Y.}\ \bibnamefont
  {Oreg}}, \bibinfo {author} {\bibfnamefont {G.}\ \bibnamefont {Refael}},\
  and\ \bibinfo {author} {\bibfnamefont {F.}\ \bibnamefont {von Oppen}},\
  }\bibfield  {title} {\bibinfo {title} {Helical liquids and Majorana bound
  states in quantum wires},\ }\href
  {https://doi.org/10.1103/PhysRevLett.105.177002} {\bibfield  {journal}
  {\bibinfo  {journal} {Phys. Rev. Lett.}\ }\textbf {\bibinfo {volume} {105}},\
  \bibinfo {pages} {177002} (\bibinfo {year} {2010})}\BibitemShut {NoStop}%
\bibitem [{\citenamefont {Mourik}\ \emph {et~al.}(2012)\citenamefont {Mourik},
  \citenamefont {Zuo}, \citenamefont {Frolov}, \citenamefont {Plissard},
  \citenamefont {Bakkers},\ and\ \citenamefont
  {Kouwenhoven}}]{Science-336-1003}%
  \BibitemOpen
  \bibfield  {author} {\bibinfo {author} {\bibfnamefont {V.}~\bibnamefont
  {Mourik}}, \bibinfo {author} {\bibfnamefont {K.}~\bibnamefont {Zuo}},
  \bibinfo {author} {\bibfnamefont {S.~M.}\ \bibnamefont {Frolov}}, \bibinfo
  {author} {\bibfnamefont {S.~R.}\ \bibnamefont {Plissard}}, \bibinfo {author}
  {\bibfnamefont {E.~P. A.~M.}\ \bibnamefont {Bakkers}},\ and\ \bibinfo
  {author} {\bibfnamefont {L.~P.}\ \bibnamefont {Kouwenhoven}},\ }\bibfield
  {title} {\bibinfo {title} {Signatures of Majorana fermions in hybrid
  superconductor-semiconductor nanowire devices},\ }\href
  {https://doi.org/10.1126/science.1222360} {\bibfield  {journal} {\bibinfo
  {journal} {Science}\ }\textbf {\bibinfo {volume} {336}},\ \bibinfo {pages}
  {1003} (\bibinfo {year} {2012})}\BibitemShut {NoStop}%
\bibitem [{\citenamefont {Deng}\ \emph {et~al.}(2012)\citenamefont {Deng},
  \citenamefont {Yu}, \citenamefont {Huang}, \citenamefont {Larsson},
  \citenamefont {Caroff},\ and\ \citenamefont {Xu}}]{NL-12-6414}%
  \BibitemOpen
  \bibfield  {author} {\bibinfo {author} {\bibfnamefont {M.~T.}\ \bibnamefont
  {Deng}}, \bibinfo {author} {\bibfnamefont {C.~L.}\ \bibnamefont {Yu}},
  \bibinfo {author} {\bibfnamefont {G.~Y.}\ \bibnamefont {Huang}}, \bibinfo
  {author} {\bibfnamefont {M.}~\bibnamefont {Larsson}}, \bibinfo {author}
  {\bibfnamefont {P.}~\bibnamefont {Caroff}},\ and\ \bibinfo {author}
  {\bibfnamefont {H.~Q.}\ \bibnamefont {Xu}},\ }\bibfield  {title} {\bibinfo
  {title} {Anomalous zero-bias conductance peak in a Nb-InSb nanowire-Nb hybrid
  device},\ }\href {https://doi.org/10.1021/nl303758w} {\bibfield  {journal}
  {\bibinfo  {journal} {Nano Lett.}\ }\textbf {\bibinfo {volume} {12}},\
  \bibinfo {pages} {6414} (\bibinfo {year} {2012})}\BibitemShut {NoStop}%
\bibitem [{\citenamefont {Das}\ \emph {et~al.}(2012)\citenamefont {Das},
  \citenamefont {Ronen}, \citenamefont {Most}, \citenamefont {Oreg},
  \citenamefont {Heiblum},\ and\ \citenamefont {Shtrikman}}]{NatPhys-8-887}%
  \BibitemOpen
  \bibfield  {author} {\bibinfo {author} {\bibfnamefont {A.}\ \bibnamefont
  {Das}}, \bibinfo {author} {\bibfnamefont {Y.}\ \bibnamefont {Ronen}},
  \bibinfo {author} {\bibfnamefont {Y.}\ \bibnamefont {Most}}, \bibinfo
  {author} {\bibfnamefont {Y.}\ \bibnamefont {Oreg}}, \bibinfo {author}
  {\bibfnamefont {M.}\ \bibnamefont {Heiblum}},\ and\ \bibinfo {author}
  {\bibfnamefont {H.}\ \bibnamefont {Shtrikman}},\ }\bibfield  {title}
  {\bibinfo {title} {Zero-bias peaks and splitting in an Al-InAs nanowire
  topological superconductor as a signature of Majorana fermions},\ }\href
  {https://doi.org/10.1038/NPHYS2479} {\bibfield  {journal} {\bibinfo
  {journal} {Nat. Phys.}\ }\textbf {\bibinfo {volume} {8}},\ \bibinfo {pages}
  {887} (\bibinfo {year} {2012})}\BibitemShut {NoStop}%
\bibitem [{\citenamefont {Rokhinson}\ \emph {et~al.}(2012)\citenamefont
  {Rokhinson}, \citenamefont {Liu},\ and\ \citenamefont
  {Furdyna}}]{NatPhys-8-795}%
  \BibitemOpen
  \bibfield  {author} {\bibinfo {author} {\bibfnamefont {L.~P.}\
  \bibnamefont {Rokhinson}}, \bibinfo {author} {\bibfnamefont {X.}\
  \bibnamefont {Liu}},\ and\ \bibinfo {author} {\bibfnamefont {J.~K.}\
  \bibnamefont {Furdyna}},\ }\bibfield  {title} {\bibinfo {title} {The
  fractional a.c. Josephson effect in a semiconductor-superconductor nanowire
  as a signature of Majorana particles},\ }\href
  {https://doi.org/10.1038/NPHYS2429} {\bibfield  {journal} {\bibinfo
  {journal} {Nat. Phys.}\ }\textbf {\bibinfo {volume} {8}},\ \bibinfo {pages}
  {795} (\bibinfo {year} {2012})}\BibitemShut {NoStop}%
\bibitem [{\citenamefont {Hassler}\ \emph {et~al.}(2010)\citenamefont
  {Hassler}, \citenamefont {Akhmerov}, \citenamefont {Hou},\ and\ \citenamefont
  {Beenakker}}]{NJP-12-125002}%
  \BibitemOpen
  \bibfield  {author} {\bibinfo {author} {\bibfnamefont {F.}~\bibnamefont
  {Hassler}}, \bibinfo {author} {\bibfnamefont {A.~R.}\ \bibnamefont {Akhmerov}},
  \bibinfo {author} {\bibfnamefont {C.-Y.}\ \bibnamefont {Hou}},\ and\ \bibinfo
  {author} {\bibfnamefont {C.~W.~J.}\ \bibnamefont {Beenakker}},\ }\bibfield
  {title} {\bibinfo {title} {Anyonic interferometry without anyons: how a flux
  qubit can read out a topological qubit},\ }\href
  {http://stacks.iop.org/1367-2630/12/i=12/a=125002} {\bibfield  {journal}
  {\bibinfo  {journal} {New J. Phys.}\ }\textbf {\bibinfo {volume} {12}},\
  \bibinfo {pages} {125002} (\bibinfo {year} {2010})}\BibitemShut {NoStop}%
\bibitem [{\citenamefont {Xue}\ \emph {et~al.}(2009)\citenamefont {Xue},
  \citenamefont {Zhu}, \citenamefont {You},\ and\ \citenamefont
  {Wang}}]{PRA-79-040303}%
  \BibitemOpen
  \bibfield  {author} {\bibinfo {author} {\bibfnamefont {Z.-Y.}\
  \bibnamefont {Xue}}, \bibinfo {author} {\bibfnamefont {S.-L.}\
  \bibnamefont {Zhu}}, \bibinfo {author} {\bibfnamefont {J.-Q.}\ \bibnamefont
  {You}},\ and\ \bibinfo {author} {\bibfnamefont {Z.-D.}\ \bibnamefont
  {Wang}},\ }\bibfield  {title} {\bibinfo {title} {Implementing topological
  quantum manipulation with superconducting circuits},\ }\href
  {https://doi.org/10.1103/PhysRevA.79.040303} {\bibfield  {journal} {\bibinfo
  {journal} {Phys. Rev. A}\ }\textbf {\bibinfo {volume} {79}},\ \bibinfo
  {pages} {040303} (\bibinfo {year} {2009})}\BibitemShut {NoStop}%
\bibitem [{\citenamefont {Flensberg}(2011)}]{PRL-106-090503}%
  \BibitemOpen
  \bibfield  {author} {\bibinfo {author} {\bibfnamefont {K.}\ \bibnamefont
  {Flensberg}},\ }\bibfield  {title} {\bibinfo {title} {Non-abelian operations
  on Majorana fermions via single-charge control},\ }\href
  {https://doi.org/10.1103/PhysRevLett.106.090503} {\bibfield  {journal}
  {\bibinfo  {journal} {Phys. Rev. Lett.}\ }\textbf {\bibinfo {volume} {106}},\
  \bibinfo {pages} {090503} (\bibinfo {year} {2011})}\BibitemShut {NoStop}%
\bibitem [{\citenamefont {Leijnse}\ and\ \citenamefont
  {Flensberg}(2011)}]{PRL-107-210502}%
  \BibitemOpen
  \bibfield  {author} {\bibinfo {author} {\bibfnamefont {M.}\ \bibnamefont
  {Leijnse}}\ and\ \bibinfo {author} {\bibfnamefont {K.}\ \bibnamefont
  {Flensberg}},\ }\bibfield  {title} {\bibinfo {title} {Quantum information
  transfer between topological and spin qubit systems},\ }\href
  {https://doi.org/10.1103/PhysRevLett.107.210502} {\bibfield  {journal}
  {\bibinfo  {journal} {Phys. Rev. Lett.}\ }\textbf {\bibinfo {volume} {107}},\
  \bibinfo {pages} {210502} (\bibinfo {year} {2011})}\BibitemShut {NoStop}%
\bibitem [{\citenamefont {Hoffman}\ \emph {et~al.}(2016)\citenamefont
  {Hoffman}, \citenamefont {Schrade}, \citenamefont {Klinovaja},\ and\
  \citenamefont {Loss}}]{PhysRevB-94-045316}%
  \BibitemOpen
  \bibfield  {author} {\bibinfo {author} {\bibfnamefont {S.}\ \bibnamefont
  {Hoffman}}, \bibinfo {author} {\bibfnamefont {C.}\ \bibnamefont
  {Schrade}}, \bibinfo {author} {\bibfnamefont {J.}\ \bibnamefont
  {Klinovaja}},\ and\ \bibinfo {author} {\bibfnamefont {D.}\ \bibnamefont
  {Loss}},\ }\bibfield  {title} {\bibinfo {title} {Universal quantum
  computation with hybrid spin-Majorana qubits},\ }\href
  {https://doi.org/10.1103/PhysRevB.94.045316} {\bibfield  {journal} {\bibinfo
  {journal} {Phys. Rev. B}\ }\textbf {\bibinfo {volume} {94}},\ \bibinfo
  {pages} {045316} (\bibinfo {year} {2016})}\BibitemShut {NoStop}%
\bibitem [{\citenamefont {Plugge}\ \emph {et~al.}(2016)\citenamefont {Plugge},
  \citenamefont {Landau}, \citenamefont {Sela}, \citenamefont {Altland},
  \citenamefont {Flensberg},\ and\ \citenamefont {Egger}}]{PhysRevB-94-174514}%
  \BibitemOpen
  \bibfield  {author} {\bibinfo {author} {\bibfnamefont {S.}~\bibnamefont
  {Plugge}}, \bibinfo {author} {\bibfnamefont {L.~A.}\ \bibnamefont {Landau}},
  \bibinfo {author} {\bibfnamefont {E.}~\bibnamefont {Sela}}, \bibinfo {author}
  {\bibfnamefont {A.}~\bibnamefont {Altland}}, \bibinfo {author} {\bibfnamefont
  {K.}~\bibnamefont {Flensberg}},\ and\ \bibinfo {author} {\bibfnamefont
  {R.}~\bibnamefont {Egger}},\ }\bibfield  {title} {\bibinfo {title} {Roadmap
  to Majorana surface codes},\ }\href
  {https://doi.org/10.1103/PhysRevB.94.174514} {\bibfield  {journal} {\bibinfo
  {journal} {Phys. Rev. B}\ }\textbf {\bibinfo {volume} {94}},\ \bibinfo
  {pages} {174514} (\bibinfo {year} {2016})}\BibitemShut {NoStop}%
\bibitem [{\citenamefont {Zhang}\ and\ \citenamefont
  {Yu}(2013)}]{PRA-87-032327}%
  \BibitemOpen
  \bibfield  {author} {\bibinfo {author} {\bibfnamefont {Z.-T.}\
  \bibnamefont {Zhang}}\ and\ \bibinfo {author} {\bibfnamefont {Y.}\
  \bibnamefont {Yu}},\ }\bibfield  {title} {\bibinfo {title} {Processing
  quantum information in a hybrid topological qubit and superconducting flux
  qubit system},\ }\href {https://doi.org/10.1103/PhysRevA.87.032327}
  {\bibfield  {journal} {\bibinfo  {journal} {Phys. Rev. A}\ }\textbf {\bibinfo
  {volume} {87}},\ \bibinfo {pages} {032327} (\bibinfo {year}
  {2013})}\BibitemShut {NoStop}%
\bibitem [{\citenamefont {Xue}\ \emph {et~al.}(2013)\citenamefont {Xue},
  \citenamefont {Shao}, \citenamefont {Hu}, \citenamefont {Zhu},\ and\
  \citenamefont {Wang}}]{PRA-88-024303}%
  \BibitemOpen
  \bibfield  {author} {\bibinfo {author} {\bibfnamefont {Z.-Y.}\
  \bibnamefont {Xue}}, \bibinfo {author} {\bibfnamefont {L.~B.}\ \bibnamefont
  {Shao}}, \bibinfo {author} {\bibfnamefont {Y.}\ \bibnamefont {Hu}},
  \bibinfo {author} {\bibfnamefont {S.-L.}\ \bibnamefont {Zhu}},\ and\
  \bibinfo {author} {\bibfnamefont {Z.~D.}\ \bibnamefont {Wang}},\ }\bibfield
  {title} {\bibinfo {title} {Tunable interfaces for realizing universal quantum
  computation with topological qubits},\ }\href
  {https://doi.org/10.1103/PhysRevA.88.024303} {\bibfield  {journal} {\bibinfo
  {journal} {Phys. Rev. A}\ }\textbf {\bibinfo {volume} {88}},\ \bibinfo
  {pages} {024303} (\bibinfo {year} {2013})}\BibitemShut {NoStop}%
\bibitem [{\citenamefont {Hong}\ \emph {et~al.}(2013)\citenamefont {Hong},
  \citenamefont {Qian}, \citenamefont {Fu}, \citenamefont {Zhu},\ and\
  \citenamefont {Jiang}}]{PRA-87-032339}%
  \BibitemOpen
  \bibfield  {author} {\bibinfo {author} {\bibfnamefont {F.-Y.}\ \bibnamefont
  {Hong}}, \bibinfo {author} {\bibfnamefont {H.-Q.}\ \bibnamefont {Qian}},
  \bibinfo {author} {\bibfnamefont {J.-L.}\ \bibnamefont {Fu}}, \bibinfo
  {author} {\bibfnamefont {Z.-Y.}\ \bibnamefont {Zhu}},\ and\ \bibinfo
  {author} {\bibfnamefont {L.-Z.}\ \bibnamefont {Jiang}},\ }\bibfield
  {title} {\bibinfo {title} {Strong coupling between a topological qubit and a
  nanomechanical resonator},\ }\href
  {https://doi.org/10.1103/PhysRevA.87.032339} {\bibfield  {journal} {\bibinfo
  {journal} {Phys. Rev. A}\ }\textbf {\bibinfo {volume} {87}},\ \bibinfo
  {pages} {032339} (\bibinfo {year} {2013})}\BibitemShut {NoStop}%
\bibitem [{\citenamefont {Jiang}\ \emph
  {et~al.}(2011{\natexlab{a}})\citenamefont {Jiang}, \citenamefont {Kane},\
  and\ \citenamefont {Preskill}}]{PRL-106-130504}%
  \BibitemOpen
  \bibfield  {author} {\bibinfo {author} {\bibfnamefont {L.}\ \bibnamefont
  {Jiang}}, \bibinfo {author} {\bibfnamefont {C.~L.}\ \bibnamefont
  {Kane}},\ and\ \bibinfo {author} {\bibfnamefont {J.}\ \bibnamefont
  {Preskill}},\ }\bibfield  {title} {\bibinfo {title} {Interface between
  topological and superconducting qubits},\ }\href
  {https://doi.org/10.1103/PhysRevLett.106.130504} {\bibfield  {journal}
  {\bibinfo  {journal} {Phys. Rev. Lett.}\ }\textbf {\bibinfo {volume} {106}},\
  \bibinfo {pages} {130504} (\bibinfo {year} {2011}{\natexlab{a}})}\BibitemShut
  {NoStop}%
\bibitem [{\citenamefont {Bonderson}\ and\ \citenamefont
  {Lutchyn}(2011)}]{PRL-106-130505}%
  \BibitemOpen
  \bibfield  {author} {\bibinfo {author} {\bibfnamefont {P.}\ \bibnamefont
  {Bonderson}}\ and\ \bibinfo {author} {\bibfnamefont {R.~M.}\ \bibnamefont
  {Lutchyn}},\ }\bibfield  {title} {\bibinfo {title} {Topological quantum
  buses: Coherent quantum information transfer between topological and
  conventional qubits},\ }\href
  {https://doi.org/10.1103/PhysRevLett.106.130505} {\bibfield  {journal}
  {\bibinfo  {journal} {Phys. Rev. Lett.}\ }\textbf {\bibinfo {volume} {106}},\
  \bibinfo {pages} {130505} (\bibinfo {year} {2011})}\BibitemShut {NoStop}%
\bibitem [{\citenamefont {Vishveshwara}(2011)}]{NP-7-450}%
  \BibitemOpen
  \bibfield  {author} {\bibinfo {author} {\bibfnamefont {S.}\ \bibnamefont
  {Vishveshwara}},\ }\bibfield  {title} {\bibinfo {title} {A bit of both},\
  }\href {https://doi.org/10.1038/nphys2014} {\bibfield  {journal} {\bibinfo
  {journal} {Nature Phys.}\ }\textbf {\bibinfo {volume} {7}},\ \bibinfo {pages}
  {450} (\bibinfo {year} {2011})}\BibitemShut {NoStop}%
\bibitem [{\citenamefont {Ohm}\ and\ \citenamefont
  {Hassler}(2014)}]{NJP-16-015009}%
  \BibitemOpen
  \bibfield  {author} {\bibinfo {author} {\bibfnamefont {C.}\
  \bibnamefont {Ohm}}\ and\ \bibinfo {author} {\bibfnamefont {F.}\
  \bibnamefont {Hassler}},\ }\bibfield  {title} {\bibinfo {title} {Majorana
  fermions coupled to electromagnetic radiation},\ }\href
  {https://doi.org/10.1088/1367-2630/16/1/015009} {\bibfield  {journal}
  {\bibinfo  {journal} {New J. Phys.}\ }\textbf {\bibinfo {volume} {16}},\
  \bibinfo {pages} {015009} (\bibinfo {year} {2014})}\BibitemShut {NoStop}%
\bibitem [{\citenamefont {Cottet}\ \emph {et~al.}(2013)\citenamefont {Cottet},
  \citenamefont {Kontos},\ and\ \citenamefont {Dou\ifmmode~\mbox{\c{c}}\else
  \c{c}\fi{}ot}}]{PRB-88-195415}%
  \BibitemOpen
  \bibfield  {author} {\bibinfo {author} {\bibfnamefont {A.}\ \bibnamefont
  {Cottet}}, \bibinfo {author} {\bibfnamefont {T.}\ \bibnamefont {Kontos}},\
  and\ \bibinfo {author} {\bibfnamefont {B.}\ \bibnamefont
  {Dou\ifmmode~\mbox{\c{c}}\else \c{c}\fi{}ot}},\ }\bibfield  {title} {\bibinfo
  {title} {Squeezing light with Majorana fermions},\ }\href
  {https://doi.org/10.1103/PhysRevB.88.195415} {\bibfield  {journal} {\bibinfo
  {journal} {Phys. Rev. B}\ }\textbf {\bibinfo {volume} {88}},\ \bibinfo
  {pages} {195415} (\bibinfo {year} {2013})}\BibitemShut {NoStop}%
\bibitem [{\citenamefont {Xue}\ \emph {et~al.}(2015)\citenamefont {Xue},
  \citenamefont {Gong}, \citenamefont {Liu}, \citenamefont {Hu}, \citenamefont
  {Zhu},\ and\ \citenamefont {Wang}}]{SR-5-12233}%
  \BibitemOpen
  \bibfield  {author} {\bibinfo {author} {\bibfnamefont {Z.-Y.}\
  \bibnamefont {Xue}}, \bibinfo {author} {\bibfnamefont {M.}\ \bibnamefont
  {Gong}}, \bibinfo {author} {\bibfnamefont {J.}\ \bibnamefont {Liu}},
  \bibinfo {author} {\bibfnamefont {Y.}\ \bibnamefont {Hu}}, \bibinfo
  {author} {\bibfnamefont {S.-L.}\ \bibnamefont {Zhu}},\ and\ \bibinfo
  {author} {\bibfnamefont {Z.~D.}\ \bibnamefont {Wang}},\ }\bibfield  {title}
  {\bibinfo {title} {Robust interface between flying and topological qubits},\
  }\href {https://doi.org/10.1038/srep12233} {\bibfield  {journal} {\bibinfo
  {journal} {Sci. Rep.}\ }\textbf {\bibinfo {volume} {5}},\ \bibinfo {pages}
  {12233} (\bibinfo {year} {2015})}\BibitemShut {NoStop}%
\bibitem [{\citenamefont {Doherty}\ \emph {et~al.}(2013)\citenamefont
  {Doherty}, \citenamefont {Manson}, \citenamefont {Delaney}, \citenamefont
  {Jelezko}, \citenamefont {Wrachtrup},\ and\ \citenamefont
  {Hollenberg}}]{PR-528-1}%
  \BibitemOpen
  \bibfield  {author} {\bibinfo {author} {\bibfnamefont {M.~W.}\
  \bibnamefont {Doherty}}, \bibinfo {author} {\bibfnamefont {N.~B.}\
  \bibnamefont {Manson}}, \bibinfo {author} {\bibfnamefont {P.}\ \bibnamefont
  {Delaney}}, \bibinfo {author} {\bibfnamefont {F.}\ \bibnamefont
  {Jelezko}}, \bibinfo {author} {\bibfnamefont {J.}\ \bibnamefont
  {Wrachtrup}},\ and\ \bibinfo {author} {\bibfnamefont {L.~C.~L.}\
  \bibnamefont {Hollenberg}},\ }\bibfield  {title} {\bibinfo {title} {The
  nitrogen-vacancy colour centre in diamond},\ }\href
  {https://doi.org/10.1016/j.physrep.2013.02.001} {\bibfield
  {journal} {\bibinfo  {journal} {Phys. Rep.}\ }\textbf {\bibinfo {volume}
  {528}},\ \bibinfo {pages} {1} (\bibinfo {year} {2013})}\BibitemShut
  {NoStop}%
\bibitem [{\citenamefont {Hong}\ \emph {et~al.}(2012)\citenamefont {Hong},
  \citenamefont {Grinolds}, \citenamefont {Maletinsky}, \citenamefont
  {Walsworth}, \citenamefont {Lukin},\ and\ \citenamefont
  {Yacoby}}]{NL-12-3920}%
  \BibitemOpen
  \bibfield  {author} {\bibinfo {author} {\bibfnamefont {S.}\ \bibnamefont
  {Hong}}, \bibinfo {author} {\bibfnamefont {M.~S.}\ \bibnamefont
  {Grinolds}}, \bibinfo {author} {\bibfnamefont {P.}\ \bibnamefont
  {Maletinsky}}, \bibinfo {author} {\bibfnamefont {R.~L.}\ \bibnamefont
  {Walsworth}}, \bibinfo {author} {\bibfnamefont {M.~D.}\ \bibnamefont
  {Lukin}},\ and\ \bibinfo {author} {\bibfnamefont {A.}\ \bibnamefont
  {Yacoby}},\ }\bibfield  {title} {\bibinfo {title} {Coherent, mechanical
  control of a single electronic spin},\ }\href
  {https://doi.org/10.1021/nl300775c} {\bibfield  {journal} {\bibinfo
  {journal} {Nano. Lett.}\ }\textbf {\bibinfo {volume} {12}},\ \bibinfo {pages}
  {3920} (\bibinfo {year} {2012})}\BibitemShut {NoStop}%
\bibitem [{\citenamefont {Bar-Gill}\ \emph {et~al.}(2013)\citenamefont
  {Bar-Gill}, \citenamefont {Pham}, \citenamefont {Jarmola}, \citenamefont
  {Budker},\ and\ \citenamefont {Walsworth}}]{NC-4-1743}%
  \BibitemOpen
  \bibfield  {author} {\bibinfo {author} {\bibfnamefont {N.}~\bibnamefont
  {Bar-Gill}}, \bibinfo {author} {\bibfnamefont {L.~M.}\ \bibnamefont {Pham}},
  \bibinfo {author} {\bibfnamefont {A.}~\bibnamefont {Jarmola}}, \bibinfo
  {author} {\bibfnamefont {D.}~\bibnamefont {Budker}},\ and\ \bibinfo {author}
  {\bibfnamefont {R.~L.}\ \bibnamefont {Walsworth}},\ }\bibfield  {title}
  {\bibinfo {title} {Solid-state electronic spin coherence time approaching one
  second},\ }\href {https://doi.org/10.1038/ncomms2771} {\bibfield  {journal}
  {\bibinfo  {journal} {Nat. Commun.}\ }\textbf {\bibinfo {volume} {4}},\
  \bibinfo {pages} {1743} (\bibinfo {year} {2013})}\BibitemShut {NoStop}%
\bibitem [{\citenamefont {Xia}\ and\ \citenamefont
  {Twamley}(2016)}]{prb-94-205118}%
  \BibitemOpen
  \bibfield  {author} {\bibinfo {author} {\bibfnamefont {K.-Y.}\ \bibnamefont
  {Xia}}\ and\ \bibinfo {author} {\bibfnamefont {J.}\ \bibnamefont
  {Twamley}},\ }\bibfield  {title} {\bibinfo {title} {Generating spin squeezing
  states and Greenberger-Horne-Zeilinger entanglement using a hybrid
  phonon-spin ensemble in diamond},\ }\href
  {https://doi.org/10.1103/PhysRevB.94.205118} {\bibfield  {journal} {\bibinfo
  {journal} {Phys. Rev. B}\ }\textbf {\bibinfo {volume} {94}},\ \bibinfo
  {pages} {205118} (\bibinfo {year} {2016})}\BibitemShut {NoStop}%
\bibitem [{\citenamefont {Yang}\ \emph
  {et~al.}(2011{\natexlab{a}})\citenamefont {Yang}, \citenamefont {Hu},
  \citenamefont {Yin}, \citenamefont {Deng},\ and\ \citenamefont
  {Feng}}]{PRA-83-022302}%
  \BibitemOpen
  \bibfield  {author} {\bibinfo {author} {\bibfnamefont {W.~L.}\ \bibnamefont
  {Yang}}, \bibinfo {author} {\bibfnamefont {Y.}~\bibnamefont {Hu}}, \bibinfo
  {author} {\bibfnamefont {Z.~Q.}\ \bibnamefont {Yin}}, \bibinfo {author}
  {\bibfnamefont {Z.~J.}\ \bibnamefont {Deng}},\ and\ \bibinfo {author}
  {\bibfnamefont {M.}~\bibnamefont {Feng}},\ }\bibfield  {title} {\bibinfo
  {title} {Entanglement of nitrogen-vacancy-center ensembles using transmission
  line resonators and a superconducting phase qubit},\ }\href
  {https://doi.org/10.1103/PhysRevA.83.022302} {\bibfield  {journal} {\bibinfo
  {journal} {Phys. Rev. A}\ }\textbf {\bibinfo {volume} {83}},\ \bibinfo
  {pages} {022302} (\bibinfo {year} {2011}{\natexlab{a}})}\BibitemShut
  {NoStop}%
\bibitem [{\citenamefont {Li}\ \emph {et~al.}(2017{\natexlab{a}})\citenamefont
  {Li}, \citenamefont {Li}, \citenamefont {Ma},\ and\ \citenamefont
  {Li}}]{SR-7-14116}%
  \BibitemOpen
  \bibfield  {author} {\bibinfo {author} {\bibfnamefont {X.-X.}\
  \bibnamefont {Li}}, \bibinfo {author} {\bibfnamefont {P.-B.}\ \bibnamefont
  {Li}}, \bibinfo {author} {\bibfnamefont {S.-L.}\ \bibnamefont {Ma}},\ and\
  \bibinfo {author} {\bibfnamefont {F.-L.}\ \bibnamefont {Li}},\ }\bibfield
  {title} {\bibinfo {title} {Preparing entangled states between two NV centers
  via the damping of nanomechanical resonators},\ }\href
  {https://doi.org/10.1038/s41598-017-14245-8} {\bibfield  {journal} {\bibinfo
  {journal} {Sci. Rep.}\ }\textbf {\bibinfo {volume} {7}},\ \bibinfo {pages}
  {14116} (\bibinfo {year} {2017}{\natexlab{a}})}\BibitemShut {NoStop}%
\bibitem [{\citenamefont {Zhou}\ \emph {et~al.}(2017)\citenamefont {Zhou},
  \citenamefont {Ma}, \citenamefont {Li}, \citenamefont {Li}, \citenamefont
  {Li},\ and\ \citenamefont {Li}}]{pra-96-062333}%
  \BibitemOpen
  \bibfield  {author} {\bibinfo {author} {\bibfnamefont {Y.}\ \bibnamefont
  {Zhou}}, \bibinfo {author} {\bibfnamefont {S.-L.}\ \bibnamefont {Ma}},
  \bibinfo {author} {\bibfnamefont {B.}~\bibnamefont {Li}}, \bibinfo {author}
  {\bibfnamefont {X.-X.}\ \bibnamefont {Li}}, \bibinfo {author}
  {\bibfnamefont {F.-L.}\ \bibnamefont {Li}},\ and\ \bibinfo {author}
  {\bibfnamefont {P.-B.}\ \bibnamefont {Li}},\ }\bibfield  {title} {\bibinfo
  {title} {Simulating the Lipkin-Meshkov-Glick model in a hybrid quantum
  system},\ }\href {https://doi.org/10.1103/PhysRevA.96.062333} {\bibfield
  {journal} {\bibinfo  {journal} {Phys. Rev. A}\ }\textbf {\bibinfo {volume}
  {96}},\ \bibinfo {pages} {062333} (\bibinfo {year} {2017})}\BibitemShut
  {NoStop}%
\bibitem [{\citenamefont {Yin}\ \emph {et~al.}(2013)\citenamefont {Yin},
  \citenamefont {Li}, \citenamefont {Zhang},\ and\ \citenamefont
  {Duan}}]{PRA-88-033614}%
  \BibitemOpen
  \bibfield  {author} {\bibinfo {author} {\bibfnamefont {Z.-Q.}\
  \bibnamefont {Yin}}, \bibinfo {author} {\bibfnamefont {T.-C.}\
  \bibnamefont {Li}}, \bibinfo {author} {\bibfnamefont {X.}\ \bibnamefont
  {Zhang}},\ and\ \bibinfo {author} {\bibfnamefont {L.-M.}\ \bibnamefont
  {Duan}},\ }\bibfield  {title} {\bibinfo {title} {Large quantum superpositions
  of a levitated nanodiamond through spin-optomechanical coupling},\ }\href
  {https://doi.org/10.1103/PhysRevA.88.033614} {\bibfield  {journal} {\bibinfo
  {journal} {Phys. Rev. A}\ }\textbf {\bibinfo {volume} {88}},\ \bibinfo
  {pages} {033614} (\bibinfo {year} {2013})}\BibitemShut {NoStop}%
\bibitem [{\citenamefont {Chen}\ \emph {et~al.}(2011)\citenamefont {Chen},
  \citenamefont {Yang}, \citenamefont {Feng},\ and\ \citenamefont
  {Du}}]{PRA-83-054305}%
  \BibitemOpen
  \bibfield  {author} {\bibinfo {author} {\bibfnamefont {Q.}\ \bibnamefont
  {Chen}}, \bibinfo {author} {\bibfnamefont {W.-L.}\ \bibnamefont {Yang}},
  \bibinfo {author} {\bibfnamefont {M.}\ \bibnamefont {Feng}},\ and\ \bibinfo
  {author} {\bibfnamefont {J.-F.}\ \bibnamefont {Du}},\ }\bibfield  {title}
  {\bibinfo {title} {Entangling separate nitrogen-vacancy centers in a scalable
  fashion via coupling to microtoroidal resonators},\ }\href
  {https://doi.org/10.1103/PhysRevA.83.054305} {\bibfield  {journal} {\bibinfo
  {journal} {Phys. Rev. A}\ }\textbf {\bibinfo {volume} {83}},\ \bibinfo
  {pages} {054305} (\bibinfo {year} {2011})}\BibitemShut {NoStop}%
\bibitem [{\citenamefont {Zu}\ \emph {et~al.}(2014)\citenamefont {Zu},
  \citenamefont {Wang}, \citenamefont {He}, \citenamefont {Zhang},
  \citenamefont {Dai}, \citenamefont {Wang},\ and\ \citenamefont
  {Duan}}]{Natrue-514-72}%
  \BibitemOpen
  \bibfield  {author} {\bibinfo {author} {\bibfnamefont {C.}~\bibnamefont
  {Zu}}, \bibinfo {author} {\bibfnamefont {W.-B.}\ \bibnamefont {Wang}},
  \bibinfo {author} {\bibfnamefont {L.}~\bibnamefont {He}}, \bibinfo {author}
  {\bibfnamefont {W.-G.}\ \bibnamefont {Zhang}}, \bibinfo {author}
  {\bibfnamefont {C.-Y.}\ \bibnamefont {Dai}}, \bibinfo {author} {\bibfnamefont
  {F.}~\bibnamefont {Wang}},\ and\ \bibinfo {author} {\bibfnamefont {L.-M.}\
  \bibnamefont {Duan}},\ }\bibfield  {title} {\bibinfo {title} {Experimental
  realization of universal geometric quantum gates with solid-state spins},\
  }\href {https://doi.org/10.1038/nature13729} {\bibfield  {journal} {\bibinfo
  {journal} {Nature}\ }\textbf {\bibinfo {volume} {514}},\ \bibinfo
  {pages} {72} (\bibinfo {year} {2014})}\BibitemShut {NoStop}%
\bibitem [{\citenamefont {Xu}\ \emph {et~al.}(2012)\citenamefont {Xu},
  \citenamefont {Wang}, \citenamefont {Duan}, \citenamefont {Huang},
  \citenamefont {Wang}, \citenamefont {Wang}, \citenamefont {Xu}, \citenamefont
  {Kong}, \citenamefont {Shi}, \citenamefont {Rong},\ and\ \citenamefont
  {Du}}]{PRL-109-070502}%
  \BibitemOpen
  \bibfield  {author} {\bibinfo {author} {\bibfnamefont {X.-K.}\
  \bibnamefont {Xu}}, \bibinfo {author} {\bibfnamefont {Z.-X.}\ \bibnamefont
  {Wang}}, \bibinfo {author} {\bibfnamefont {C.-K.}\ \bibnamefont {Duan}},
  \bibinfo {author} {\bibfnamefont {P.}~\bibnamefont {Huang}}, \bibinfo
  {author} {\bibfnamefont {P.-F.}\ \bibnamefont {Wang}}, \bibinfo {author}
  {\bibfnamefont {Y.}~\bibnamefont {Wang}}, \bibinfo {author} {\bibfnamefont
  {N.-Y.}\ \bibnamefont {Xu}}, \bibinfo {author} {\bibfnamefont
  {X.}~\bibnamefont {Kong}}, \bibinfo {author} {\bibfnamefont {F.-Z.}\
  \bibnamefont {Shi}}, \bibinfo {author} {\bibfnamefont {X.}\ \bibnamefont
  {Rong}},\ and\ \bibinfo {author} {\bibfnamefont {J.-F.}\ \bibnamefont
  {Du}},\ }\bibfield  {title} {\bibinfo {title} {Coherence-protected quantum
  gate by continuous dynamical decoupling in diamond},\ }\href
  {https://doi.org/10.1103/PhysRevLett.109.070502} {\bibfield  {journal}
  {\bibinfo  {journal} {Phys. Rev. Lett.}\ }\textbf {\bibinfo {volume} {109}},\
  \bibinfo {pages} {070502} (\bibinfo {year} {2012})}\BibitemShut {NoStop}%
\bibitem [{\citenamefont {Dong}\ \emph {et~al.}(2017)\citenamefont {Dong},
  \citenamefont {Rong}, \citenamefont {Geng}, \citenamefont {Shi},
  \citenamefont {Li}, \citenamefont {Duan},\ and\ \citenamefont
  {Du}}]{PRB-96-205149}%
  \BibitemOpen
  \bibfield  {author} {\bibinfo {author} {\bibfnamefont {L.-H.}\ \bibnamefont
  {Dong}}, \bibinfo {author} {\bibfnamefont {X.}\ \bibnamefont {Rong}},
  \bibinfo {author} {\bibfnamefont {J.-P.}\ \bibnamefont {Geng}}, \bibinfo
  {author} {\bibfnamefont {F.-Z.}\ \bibnamefont {Shi}}, \bibinfo {author}
  {\bibfnamefont {Z.-K.}\ \bibnamefont {Li}}, \bibinfo {author}
  {\bibfnamefont {C.-K.}\ \bibnamefont {Duan}},\ and\ \bibinfo {author}
  {\bibfnamefont {J.-F.}\ \bibnamefont {Du}},\ }\bibfield  {title}
  {\bibinfo {title} {Scalable quantum computation scheme based on
  quantum-actuated nuclear-spin decoherence-free qubits},\ }\href
  {https://doi.org/10.1103/PhysRevB.96.205149} {\bibfield  {journal} {\bibinfo
  {journal} {Phys. Rev. B}\ }\textbf {\bibinfo {volume} {96}},\ \bibinfo
  {pages} {205149} (\bibinfo {year} {2017})}\BibitemShut {NoStop}%
\bibitem [{\citenamefont {Cao}\ \emph {et~al.}(2017)\citenamefont {Cao},
  \citenamefont {Betzholz}, \citenamefont {Zhang},\ and\ \citenamefont
  {Cai}}]{prb-96-245418}%
  \BibitemOpen
  \bibfield  {author} {\bibinfo {author} {\bibfnamefont {P.-H.}\ \bibnamefont
  {Cao}}, \bibinfo {author} {\bibfnamefont {R.}\ \bibnamefont {Betzholz}},
  \bibinfo {author} {\bibfnamefont {S.-L.}\ \bibnamefont {Zhang}},\ and\
  \bibinfo {author} {\bibfnamefont {J.-M.}\ \bibnamefont {Cai}},\ }\bibfield
   {title} {\bibinfo {title} {Entangling distant solid-state spins via thermal
  phonons},\ }\href {https://doi.org/10.1103/PhysRevB.96.245418} {\bibfield
  {journal} {\bibinfo  {journal} {Phys. Rev. B}\ }\textbf {\bibinfo {volume}
  {96}},\ \bibinfo {pages} {245418} (\bibinfo {year} {2017})}\BibitemShut
  {NoStop}%
\bibitem [{\citenamefont {Li}\ \emph {et~al.}(2018)\citenamefont {Li},
  \citenamefont {Miranowicz}, \citenamefont {Hu}, \citenamefont {Xia},\ and\
  \citenamefont {Nori}}]{pra-97-062318}%
  \BibitemOpen
  \bibfield  {author} {\bibinfo {author} {\bibfnamefont {T.}\ \bibnamefont
  {Li}}, \bibinfo {author} {\bibfnamefont {A.}\ \bibnamefont {Miranowicz}},
  \bibinfo {author} {\bibfnamefont {X.-D.}\ \bibnamefont {Hu}}, \bibinfo
  {author} {\bibfnamefont {K.-Y.}\ \bibnamefont {Xia}},\ and\ \bibinfo {author}
  {\bibfnamefont {F.}\ \bibnamefont {Nori}},\ }\bibfield  {title} {\bibinfo
  {title} {Quantum memory and gates using a
  $\mathrm{\ensuremath{\Lambda}}$-type quantum emitter coupled to a chiral
  waveguide},\ }\href {https://doi.org/10.1103/PhysRevA.97.062318} {\bibfield
  {journal} {\bibinfo  {journal} {Phys. Rev. A}\ }\textbf {\bibinfo {volume}
  {97}},\ \bibinfo {pages} {062318} (\bibinfo {year} {2018})}\BibitemShut
  {NoStop}%
\bibitem [{\citenamefont {Xia}\ and\ \citenamefont
  {Twamley}(2015)}]{pra-91-042307}%
  \BibitemOpen
  \bibfield  {author} {\bibinfo {author} {\bibfnamefont {K.-Y.}\ \bibnamefont
  {Xia}}\ and\ \bibinfo {author} {\bibfnamefont {J.}\ \bibnamefont
  {Twamley}},\ }\bibfield  {title} {\bibinfo {title} {Solid-state optical
  interconnect between distant superconducting quantum chips},\ }\href
  {https://doi.org/10.1103/PhysRevA.91.042307} {\bibfield  {journal} {\bibinfo
  {journal} {Phys. Rev. A}\ }\textbf {\bibinfo {volume} {91}},\ \bibinfo
  {pages} {042307} (\bibinfo {year} {2015})}\BibitemShut {NoStop}%
\bibitem [{\citenamefont {Yang}\ \emph
  {et~al.}(2011{\natexlab{b}})\citenamefont {Yang}, \citenamefont {Yin},
  \citenamefont {Hu}, \citenamefont {Feng},\ and\ \citenamefont
  {Du}}]{PRA-84-010301}%
  \BibitemOpen
  \bibfield  {author} {\bibinfo {author} {\bibfnamefont {W.~L.}\ \bibnamefont
  {Yang}}, \bibinfo {author} {\bibfnamefont {Z.~Q.}\ \bibnamefont {Yin}},
  \bibinfo {author} {\bibfnamefont {Y.}~\bibnamefont {Hu}}, \bibinfo {author}
  {\bibfnamefont {M.}~\bibnamefont {Feng}},\ and\ \bibinfo {author}
  {\bibfnamefont {J.~F.}\ \bibnamefont {Du}},\ }\bibfield  {title} {\bibinfo
  {title} {High-fidelity quantum memory using nitrogen-vacancy center ensemble
  for hybrid quantum computation},\ }\href
  {https://doi.org/10.1103/PhysRevA.84.010301} {\bibfield  {journal} {\bibinfo
  {journal} {Phys. Rev. A}\ }\textbf {\bibinfo {volume} {84}},\ \bibinfo
  {pages} {010301} (\bibinfo {year} {2011}{\natexlab{b}})}\BibitemShut
  {NoStop}%
\bibitem [{\citenamefont {L\"u}\ \emph {et~al.}(2013)\citenamefont {L\"u},
  \citenamefont {Xiang}, \citenamefont {Cui}, \citenamefont {You},\ and\
  \citenamefont {Nori}}]{pra-88-012329}%
  \BibitemOpen
  \bibfield  {author} {\bibinfo {author} {\bibfnamefont {X.-Y.}\ \bibnamefont
  {L\"u}}, \bibinfo {author} {\bibfnamefont {Z.-L.}\ \bibnamefont {Xiang}},
  \bibinfo {author} {\bibfnamefont {W.}\ \bibnamefont {Cui}}, \bibinfo
  {author} {\bibfnamefont {J.~Q.}\ \bibnamefont {You}},\ and\ \bibinfo {author}
  {\bibfnamefont {F.}\ \bibnamefont {Nori}},\ }\bibfield  {title} {\bibinfo
  {title} {Quantum memory using a hybrid circuit with flux qubits and
  nitrogen-vacancy centers},\ }\href
  {https://doi.org/10.1103/PhysRevA.88.012329} {\bibfield  {journal} {\bibinfo
  {journal} {Phys. Rev. A}\ }\textbf {\bibinfo {volume} {88}},\ \bibinfo
  {pages} {012329} (\bibinfo {year} {2013})}\BibitemShut {NoStop}%
\bibitem [{\citenamefont {Li}\ \emph {et~al.}(2017{\natexlab{b}})\citenamefont
  {Li}, \citenamefont {Li}, \citenamefont {Zhou}, \citenamefont {Ma},\ and\
  \citenamefont {Li}}]{PRA-96-032342}%
  \BibitemOpen
  \bibfield  {author} {\bibinfo {author} {\bibfnamefont {B.}~\bibnamefont
  {Li}}, \bibinfo {author} {\bibfnamefont {P.-B.}\ \bibnamefont {Li}},
  \bibinfo {author} {\bibfnamefont {Y.}\ \bibnamefont {Zhou}}, \bibinfo
  {author} {\bibfnamefont {S.-L.}\ \bibnamefont {Ma}},\ and\ \bibinfo
  {author} {\bibfnamefont {F.-L.}\ \bibnamefont {Li}},\ }\bibfield  {title}
  {\bibinfo {title} {Quantum microwave-optical interface with nitrogen-vacancy
  centers in diamond},\ }\href {https://doi.org/10.1103/PhysRevA.96.032342}
  {\bibfield  {journal} {\bibinfo  {journal} {Phys. Rev. A}\ }\textbf {\bibinfo
  {volume} {96}},\ \bibinfo {pages} {032342} (\bibinfo {year}
  {2017}{\natexlab{b}})}\BibitemShut {NoStop}%
\bibitem [{\citenamefont {Xiang}\ \emph {et~al.}(2013)\citenamefont {Xiang},
  \citenamefont {L\"u}, \citenamefont {Li}, \citenamefont {You},\ and\
  \citenamefont {Nori}}]{PRB-87-144516}%
  \BibitemOpen
  \bibfield  {author} {\bibinfo {author} {\bibfnamefont {Z.-L.}\
  \bibnamefont {Xiang}}, \bibinfo {author} {\bibfnamefont {X.-Y.}\
  \bibnamefont {L\"u}}, \bibinfo {author} {\bibfnamefont {T.-F.}\ \bibnamefont
  {Li}}, \bibinfo {author} {\bibfnamefont {J.~Q.}\ \bibnamefont {You}},\ and\
  \bibinfo {author} {\bibfnamefont {F.}\ \bibnamefont {Nori}},\ }\bibfield
  {title} {\bibinfo {title} {Hybrid quantum circuit consisting of a
  superconducting flux qubit coupled to a spin ensemble and a transmission-line
  resonator},\ }\href {https://doi.org/10.1103/PhysRevB.87.144516} {\bibfield
  {journal} {\bibinfo  {journal} {Phys. Rev. B}\ }\textbf {\bibinfo {volume}
  {87}},\ \bibinfo {pages} {144516} (\bibinfo {year} {2013})}\BibitemShut
  {NoStop}%
\bibitem [{\citenamefont {You}\ \emph {et~al.}(2014)\citenamefont {You},
  \citenamefont {Yang}, \citenamefont {Xu}, \citenamefont {Chan},\ and\
  \citenamefont {Oh}}]{PRB-90-195112}%
  \BibitemOpen
  \bibfield  {author} {\bibinfo {author} {\bibfnamefont {J.-B.}\ \bibnamefont
  {You}}, \bibinfo {author} {\bibfnamefont {W.-L.}\ \bibnamefont {Yang}},
  \bibinfo {author} {\bibfnamefont {Z.-Y.}\ \bibnamefont {Xu}}, \bibinfo
  {author} {\bibfnamefont {A.~H.}\ \bibnamefont {Chan}},\ and\ \bibinfo
  {author} {\bibfnamefont {C.~H.}\ \bibnamefont {Oh}},\ }\bibfield  {title}
  {\bibinfo {title} {Phase transition of light in circuit-QED lattices coupled
  to nitrogen-vacancy centers in diamond},\ }\href
  {https://doi.org/10.1103/PhysRevB.90.195112} {\bibfield  {journal} {\bibinfo
  {journal} {Phys. Rev. B}\ }\textbf {\bibinfo {volume} {90}},\ \bibinfo
  {pages} {195112} (\bibinfo {year} {2014})}\BibitemShut {NoStop}%
\bibitem [{\citenamefont {Kovalev}\ \emph {et~al.}(2014)\citenamefont
  {Kovalev}, \citenamefont {De},\ and\ \citenamefont
  {Shtengel}}]{PRL-112-106402}%
  \BibitemOpen
  \bibfield  {author} {\bibinfo {author} {\bibfnamefont {A.~A.}\
  \bibnamefont {Kovalev}}, \bibinfo {author} {\bibfnamefont {A.}\
  \bibnamefont {De}},\ and\ \bibinfo {author} {\bibfnamefont {K.}\
  \bibnamefont {Shtengel}},\ }\bibfield  {title} {\bibinfo {title} {Spin
  transfer of quantum information between Majorana modes and a resonator},\
  }\href {https://doi.org/10.1103/PhysRevLett.112.106402} {\bibfield  {journal}
  {\bibinfo  {journal} {Phys. Rev. Lett.}\ }\textbf {\bibinfo {volume} {112}},\
  \bibinfo {pages} {106402} (\bibinfo {year} {2014})}\BibitemShut {NoStop}%
\bibitem [{\citenamefont {Kastrup}(2006)}]{PRA-73-052104}%
  \BibitemOpen
  \bibfield  {author} {\bibinfo {author} {\bibfnamefont {H.~A.}\ \bibnamefont
  {Kastrup}},\ }\bibfield  {title} {\bibinfo {title} {Quantization of the
  canonically conjugate pair angle and orbital angular momentum},\ }\href
  {https://doi.org/10.1103/PhysRevA.73.052104} {\bibfield  {journal} {\bibinfo
  {journal} {Phys. Rev. A}\ }\textbf {\bibinfo {volume} {73}},\ \bibinfo
  {pages} {052104} (\bibinfo {year} {2006})}\BibitemShut {NoStop}%
\bibitem [{\citenamefont {Jiang}\ \emph {et~al.}(2013)\citenamefont {Jiang},
  \citenamefont {Pekker}, \citenamefont {Alicea}, \citenamefont {Refael},
  \citenamefont {Oreg}, \citenamefont {Brataas},\ and\ \citenamefont {von
  Oppen}}]{PRB-87-075438}%
  \BibitemOpen
  \bibfield  {author} {\bibinfo {author} {\bibfnamefont {L.}\ \bibnamefont
  {Jiang}}, \bibinfo {author} {\bibfnamefont {D.}\ \bibnamefont {Pekker}},
  \bibinfo {author} {\bibfnamefont {J.}\ \bibnamefont {Alicea}}, \bibinfo
  {author} {\bibfnamefont {G.}\ \bibnamefont {Refael}}, \bibinfo {author}
  {\bibfnamefont {Y.}\ \bibnamefont {Oreg}}, \bibinfo {author}
  {\bibfnamefont {A.}\ \bibnamefont {Brataas}},\ and\ \bibinfo {author}
  {\bibfnamefont {F.}\ \bibnamefont {von Oppen}},\ }\bibfield  {title}
  {\bibinfo {title} {Magneto-Josephson effects in junctions with Majorana bound
  states},\ }\href {https://doi.org/10.1103/PhysRevB.87.075438} {\bibfield
  {journal} {\bibinfo  {journal} {Phys. Rev. B}\ }\textbf {\bibinfo {volume}
  {87}},\ \bibinfo {pages} {075438} (\bibinfo {year} {2013})}\BibitemShut
  {NoStop}%
\bibitem [{\citenamefont {Jiang}\ \emph
  {et~al.}(2011{\natexlab{b}})\citenamefont {Jiang}, \citenamefont {Pekker},
  \citenamefont {Alicea}, \citenamefont {Refael}, \citenamefont {Oreg},\ and\
  \citenamefont {von Oppen}}]{PRL-107-236401}%
  \BibitemOpen
  \bibfield  {author} {\bibinfo {author} {\bibfnamefont {L.}\ \bibnamefont
  {Jiang}}, \bibinfo {author} {\bibfnamefont {D.}\ \bibnamefont {Pekker}},
  \bibinfo {author} {\bibfnamefont {J.}\ \bibnamefont {Alicea}}, \bibinfo
  {author} {\bibfnamefont {G.}\ \bibnamefont {Refael}}, \bibinfo {author}
  {\bibfnamefont {Y.}\ \bibnamefont {Oreg}},\ and\ \bibinfo {author}
  {\bibfnamefont {F.}\ \bibnamefont {von Oppen}},\ }\bibfield  {title}
  {\bibinfo {title} {Unconventional Josephson signatures of Majorana bound
  states},\ }\href {https://doi.org/10.1103/PhysRevLett.107.236401} {\bibfield
  {journal} {\bibinfo  {journal} {Phys. Rev. Lett.}\ }\textbf {\bibinfo
  {volume} {107}},\ \bibinfo {pages} {236401} (\bibinfo {year}
  {2011}{\natexlab{b}})}\BibitemShut {NoStop}%
\bibitem [{\citenamefont {Kotetes}\ \emph {et~al.}(2013)\citenamefont
  {Kotetes}, \citenamefont {Sch\"on},\ and\ \citenamefont
  {Shnirman}}]{JKPS-62-1558}%
  \BibitemOpen
  \bibfield  {author} {\bibinfo {author} {\bibfnamefont {P.}\
  \bibnamefont {Kotetes}}, \bibinfo {author} {\bibfnamefont {G.}\
  \bibnamefont {Sch\"on}},\ and\ \bibinfo {author} {\bibfnamefont {A.}\
  \bibnamefont {Shnirman}},\ }\bibfield  {title} {\bibinfo {title} {Engineering
  and manipulating topological qubits in 1D quantum wires},\ }\href
  {https://doi.org/10.3938/jkps.62.1558} {\bibfield  {journal} {\bibinfo
  {journal} {J. Korean Phys. Soc.}\ }\textbf {\bibinfo {volume} {62}},\
  \bibinfo {pages} {1558} (\bibinfo {year} {2013})}\BibitemShut {NoStop}%
\bibitem [{\citenamefont {Alicea}\ \emph {et~al.}(2011)\citenamefont {Alicea},
  \citenamefont {Oreg}, \citenamefont {Refael}, \citenamefont {Oppen},\ and\
  \citenamefont {Fisher}}]{NP-7-412}%
  \BibitemOpen
  \bibfield  {author} {\bibinfo {author} {\bibfnamefont {J.}\ \bibnamefont
  {Alicea}}, \bibinfo {author} {\bibfnamefont {Y.}\ \bibnamefont {Oreg}},
  \bibinfo {author} {\bibfnamefont {G.}\ \bibnamefont {Refael}}, \bibinfo
  {author} {\bibfnamefont {F.}\ \bibnamefont {von Oppen}},\ and\ \bibinfo
  {author} {\bibfnamefont {M.~P.~A.}\ \bibnamefont {Fisher}},\ }\bibfield
  {title} {\bibinfo {title} {Non-abelian statistics and topological quantum
  information processing in 1D wire networks},\ }\href
  {https://doi.org/10.1038/NPHYS1915} {\bibfield  {journal} {\bibinfo
  {journal} {Nat. Phys.}\ }\textbf {\bibinfo {volume} {7}},\ \bibinfo {pages}
  {412} (\bibinfo {year} {2011})}\BibitemShut {NoStop}%
\bibitem [{\citenamefont {Vijay}\ \emph {et~al.}(2015)\citenamefont {Vijay},
  \citenamefont {Hsieh},\ and\ \citenamefont {Fu}}]{PRX-5-041038}%
  \BibitemOpen
  \bibfield  {author} {\bibinfo {author} {\bibfnamefont {S.}\ \bibnamefont
  {Vijay}}, \bibinfo {author} {\bibfnamefont {T.~H.}\ \bibnamefont
  {Hsieh}},\ and\ \bibinfo {author} {\bibfnamefont {L.}\ \bibnamefont
  {Fu}},\ }\bibfield  {title} {\bibinfo {title} {Majorana fermion surface code
  for universal quantum computation},\ }\href
  {https://doi.org/10.1103/PhysRevX.5.041038} {\bibfield  {journal} {\bibinfo
  {journal} {Phys. Rev. X}\ }\textbf {\bibinfo {volume} {5}},\ \bibinfo {pages}
  {041038} (\bibinfo {year} {2015})}\BibitemShut {NoStop}%
\bibitem [{\citenamefont {Saket}\ \emph {et~al.}(2010)\citenamefont {Saket},
  \citenamefont {Hassan},\ and\ \citenamefont {Shankar}}]{PRB-82-174409}%
  \BibitemOpen
  \bibfield  {author} {\bibinfo {author} {\bibfnamefont {A.}\ \bibnamefont
  {Saket}}, \bibinfo {author} {\bibfnamefont {S.~R.}\ \bibnamefont {Hassan}},\
  and\ \bibinfo {author} {\bibfnamefont {R.}~\bibnamefont {Shankar}},\
  }\bibfield  {title} {\bibinfo {title} {Manipulating unpaired Majorana
  fermions in a quantum spin chain},\ }\href
  {https://doi.org/10.1103/PhysRevB.82.174409} {\bibfield  {journal} {\bibinfo
  {journal} {Phys. Rev. B}\ }\textbf {\bibinfo {volume} {82}},\ \bibinfo
  {pages} {174409} (\bibinfo {year} {2010})}\BibitemShut {NoStop}%
\bibitem [{\citenamefont {Liu}\ and\ \citenamefont
  {Drummond}(2012)}]{PRA-86-035602}%
  \BibitemOpen
  \bibfield  {author} {\bibinfo {author} {\bibfnamefont {X.-J.}\ \bibnamefont
  {Liu}}\ and\ \bibinfo {author} {\bibfnamefont {P.~D.}\ \bibnamefont
  {Drummond}},\ }\bibfield  {title} {\bibinfo {title} {Manipulating Majorana
  fermions in one-dimensional spin-orbit-coupled atomic Fermi gases},\ }\href
  {https://doi.org/10.1103/PhysRevA.86.035602} {\bibfield  {journal} {\bibinfo
  {journal} {Phys. Rev. A}\ }\textbf {\bibinfo {volume} {86}},\ \bibinfo
  {pages} {035602} (\bibinfo {year} {2012})}\BibitemShut {NoStop}%
\bibitem [{\citenamefont {Liu}\ and\ \citenamefont
  {Lobos}(2013)}]{PRB-87-060504}%
  \BibitemOpen
  \bibfield  {author} {\bibinfo {author} {\bibfnamefont {X.-J.}\
  \bibnamefont {Liu}}\ and\ \bibinfo {author} {\bibfnamefont {A.~M.}\
  \bibnamefont {Lobos}},\ }\bibfield  {title} {\bibinfo {title} {Manipulating
  Majorana fermions in quantum nanowires with broken inversion symmetry},\
  }\href {https://doi.org/10.1103/PhysRevB.87.060504} {\bibfield  {journal}
  {\bibinfo  {journal} {Phys. Rev. B}\ }\textbf {\bibinfo {volume} {87}},\
  \bibinfo {pages} {060504} (\bibinfo {year} {2013})}\BibitemShut {NoStop}%
\bibitem [{\citenamefont {Sticlet}\ \emph {et~al.}(2013)\citenamefont
  {Sticlet}, \citenamefont {Bena},\ and\ \citenamefont
  {Simon}}]{PRB-87-104509}%
  \BibitemOpen
  \bibfield  {author} {\bibinfo {author} {\bibfnamefont {D.}\ \bibnamefont
  {Sticlet}}, \bibinfo {author} {\bibfnamefont {C.}\ \bibnamefont
  {Bena}},\ and\ \bibinfo {author} {\bibfnamefont {P.}~\bibnamefont {Simon}},\
  }\bibfield  {title} {\bibinfo {title} {Josephson effect in superconducting
  wires supporting multiple Majorana edge states},\ }\href
  {https://doi.org/10.1103/PhysRevB.87.104509} {\bibfield  {journal} {\bibinfo
  {journal} {Phys. Rev. B}\ }\textbf {\bibinfo {volume} {87}},\ \bibinfo
  {pages} {104509} (\bibinfo {year} {2013})}\BibitemShut {NoStop}%
\bibitem [{\citenamefont {Kwon}\ \emph {et~al.}(2004)\citenamefont {Kwon},
  \citenamefont {Sengupta},\ and\ \citenamefont {Yakovenko}}]{EPJB-37-349}%
  \BibitemOpen
  \bibfield  {author} {\bibinfo {author} {\bibfnamefont {H.-J.}\ \bibnamefont
  {Kwon}}, \bibinfo {author} {\bibfnamefont {K.}~\bibnamefont {Sengupta}},\
  and\ \bibinfo {author} {\bibfnamefont {V.~M.}\ \bibnamefont {Yakovenko}},\
  }\bibfield  {title} {\bibinfo {title} {Fractional ac Josephson effect in p-
  and d-wave superconductors},\ }\href
  {https://doi.org/10.1140/epjb/e2004-00066-4} {\bibfield  {journal} {\bibinfo
  {journal} {Eur. Phys. J. B}\ }\textbf {\bibinfo {volume} {37}},\ \bibinfo
  {pages} {349} (\bibinfo {year} {2004})}\BibitemShut {NoStop}%
\bibitem [{\citenamefont {Sun}\ and\ \citenamefont
  {Liu}(2018)}]{PRB-97-035311}%
  \BibitemOpen
  \bibfield  {author} {\bibinfo {author} {\bibfnamefont {D.-H.}\ \bibnamefont
  {Sun}}\ and\ \bibinfo {author} {\bibfnamefont {J.}\ \bibnamefont {Liu}},\
  }\bibfield  {title} {\bibinfo {title} {Quench dynamics of the Josephson
  current in a topological Josephson junction},\ }\href
  {https://doi.org/10.1103/PhysRevB.97.035311} {\bibfield  {journal} {\bibinfo
  {journal} {Phys. Rev. B}\ }\textbf {\bibinfo {volume} {97}},\ \bibinfo
  {pages} {035311} (\bibinfo {year} {2018})}\BibitemShut {NoStop}%
\bibitem [{\citenamefont {Das~Sarma}\ \emph {et~al.}(2012)\citenamefont
  {Das~Sarma}, \citenamefont {Sau},\ and\ \citenamefont
  {Stanescu}}]{PRB-86-220506}%
  \BibitemOpen
  \bibfield  {author} {\bibinfo {author} {\bibfnamefont {S.}~\bibnamefont
  {Das~Sarma}}, \bibinfo {author} {\bibfnamefont {J.~D.}\ \bibnamefont
  {Sau}},\ and\ \bibinfo {author} {\bibfnamefont {T.~D.}\ \bibnamefont
  {Stanescu}},\ }\bibfield  {title} {\bibinfo {title} {Splitting of the
  zero-bias conductance peak as smoking gun evidence for the existence of the
  Majorana mode in a superconductor semiconductor nanowire},\ }\href
  {https://doi.org/10.1103/PhysRevB.86.220506} {\bibfield  {journal} {\bibinfo
  {journal} {Phys. Rev. B}\ }\textbf {\bibinfo {volume} {86}},\ \bibinfo
  {pages} {220506} (\bibinfo {year} {2012})}\BibitemShut {NoStop}%
\bibitem [{\citenamefont {Pientka}\ \emph {et~al.}(2013)\citenamefont
  {Pientka}, \citenamefont {Jiang}, \citenamefont {Pekker}, \citenamefont
  {Alicea}, \citenamefont {Refael}, \citenamefont {Oreg},\ and\ \citenamefont
  {von Oppen}}]{NJP-15-015502}%
  \BibitemOpen
  \bibfield  {author} {\bibinfo {author} {\bibfnamefont {F.}\ \bibnamefont
  {Pientka}}, \bibinfo {author} {\bibfnamefont {L.}\ \bibnamefont {Jiang}},
  \bibinfo {author} {\bibfnamefont {D.}\ \bibnamefont {Pekker}}, \bibinfo
  {author} {\bibfnamefont {J.}\ \bibnamefont {Alicea}}, \bibinfo {author}
  {\bibfnamefont {G.}\ \bibnamefont {Refael}}, \bibinfo {author}
  {\bibfnamefont {Y.}\ \bibnamefont {Oreg}},\ and\ \bibinfo {author}
  {\bibfnamefont {F.}\ \bibnamefont {von Oppen}},\ }\bibfield  {title}
  {\bibinfo {title} {Magneto-Josephson effects and Majorana bound states in
  quantum wires},\ }\href {https://doi.org/10.1088/1367-2630/15/11/115001}
  {\bibfield  {journal} {\bibinfo  {journal} {New J. Phys.}\ }\textbf {\bibinfo
  {volume} {15}},\ \bibinfo {pages} {115001} (\bibinfo {year}
  {2013})}\BibitemShut {NoStop}%
\bibitem [{\citenamefont {Yin}\ \emph {et~al.}(2015)\citenamefont {Yin},
  \citenamefont {Zhao},\ and\ \citenamefont {Li}}]{SCP-56-050303}%
  \BibitemOpen
  \bibfield  {author} {\bibinfo {author} {\bibfnamefont {Z.-Q.}\
  \bibnamefont {Yin}}, \bibinfo {author} {\bibfnamefont {N.}\ \bibnamefont
  {Zhao}},\ and\ \bibinfo {author} {\bibfnamefont {T.-C.}\ \bibnamefont
  {Li}},\ }\bibfield  {title} {\bibinfo {title} {Hybrid opto-mechanical systems
  with nitrogen-vacancy centers},\ }\href
  {https://doi.org/10.1007/s11433-015-5651-1} {\bibfield  {journal} {\bibinfo
  {journal} {Sci. China: Phys., Mech. Astron.}\ }\textbf {\bibinfo {volume}
  {58}},\ \bibinfo {pages} {1} (\bibinfo {year} {2015})}\BibitemShut
  {NoStop}%
\bibitem [{\citenamefont {Li}\ and\ \citenamefont
  {Nori}(2018)}]{PRApl-10-024011}%
  \BibitemOpen
  \bibfield  {author} {\bibinfo {author} {\bibfnamefont {P.-B.}\ \bibnamefont
  {Li}}\ and\ \bibinfo {author} {\bibfnamefont {F.}\ \bibnamefont {Nori}},\
  }\bibfield  {title} {\bibinfo {title} {Hybrid quantum system with
  nitrogen-vacancy centers in diamond coupled to surface-phonon polaritons in
  piezomagnetic superlattices},\ }\href
  {https://doi.org/10.1103/PhysRevApplied.10.024011} {\bibfield  {journal}
  {\bibinfo  {journal} {Phys. Rev. Appl.}\ }\textbf {\bibinfo {volume}
  {10}},\ \bibinfo {pages} {024011} (\bibinfo {year} {2018})}\BibitemShut
  {NoStop}%
\bibitem [{\citenamefont {Rabl}\ \emph {et~al.}(2009)\citenamefont {Rabl},
  \citenamefont {Cappellaro}, \citenamefont {Dutt}, \citenamefont {Jiang},
  \citenamefont {Maze},\ and\ \citenamefont {Lukin}}]{PRB-79-041302}%
  \BibitemOpen
  \bibfield  {author} {\bibinfo {author} {\bibfnamefont {P.}~\bibnamefont
  {Rabl}}, \bibinfo {author} {\bibfnamefont {P.}~\bibnamefont {Cappellaro}},
  \bibinfo {author} {\bibfnamefont {M.~V.~Gurudev}\ \bibnamefont {Dutt}},
  \bibinfo {author} {\bibfnamefont {L.}~\bibnamefont {Jiang}}, \bibinfo
  {author} {\bibfnamefont {J.~R.}\ \bibnamefont {Maze}},\ and\ \bibinfo
  {author} {\bibfnamefont {M.~D.}\ \bibnamefont {Lukin}},\ }\bibfield  {title}
  {\bibinfo {title} {Strong magnetic coupling between an electronic spin qubit
  and a mechanical resonator},\ }\href
  {https://doi.org/10.1103/PhysRevB.79.041302} {\bibfield  {journal} {\bibinfo
  {journal} {Phys. Rev. B}\ }\textbf {\bibinfo {volume} {79}},\ \bibinfo
  {pages} {041302} (\bibinfo {year} {2009})}\BibitemShut {NoStop}%
\bibitem [{\citenamefont {Ma}\ \emph {et~al.}(2017)\citenamefont {Ma},
  \citenamefont {Hoang}, \citenamefont {Gong}, \citenamefont {Li},\ and\
  \citenamefont {Yin}}]{PRA-96-023827}%
  \BibitemOpen
  \bibfield  {author} {\bibinfo {author} {\bibfnamefont {Y.}\ \bibnamefont
  {Ma}}, \bibinfo {author} {\bibfnamefont {T.~M.}\ \bibnamefont {Hoang}},
  \bibinfo {author} {\bibfnamefont {M.}\ \bibnamefont {Gong}}, \bibinfo
  {author} {\bibfnamefont {T.-C.}\ \bibnamefont {Li}},\ and\ \bibinfo
  {author} {\bibfnamefont {Z.-Q.}\ \bibnamefont {Yin}},\ }\bibfield  {title}
  {\bibinfo {title} {Proposal for quantum many-body simulation and torsional
  matter-wave interferometry with a levitated nanodiamond},\ }\href
  {https://doi.org/10.1103/PhysRevA.96.023827} {\bibfield  {journal} {\bibinfo
  {journal} {Phys. Rev. A}\ }\textbf {\bibinfo {volume} {96}},\ \bibinfo
  {pages} {023827} (\bibinfo {year} {2017})}\BibitemShut {NoStop}%
\bibitem [{\citenamefont {Delord}\ \emph {et~al.}(2017)\citenamefont {Delord},
  \citenamefont {Nicolas}, \citenamefont {Chassagneux},\ and\ \citenamefont
  {H\'etet}}]{PRA-96-063810}%
  \BibitemOpen
  \bibfield  {author} {\bibinfo {author} {\bibfnamefont {T.}~\bibnamefont
  {Delord}}, \bibinfo {author} {\bibfnamefont {L.}~\bibnamefont {Nicolas}},
  \bibinfo {author} {\bibfnamefont {Y.}~\bibnamefont {Chassagneux}},\ and\
  \bibinfo {author} {\bibfnamefont {G.}~\bibnamefont {H\'etet}},\ }\bibfield
  {title} {\bibinfo {title} {Strong coupling between a single nitrogen-vacancy
  spin and the rotational mode of diamonds levitating in an ion trap},\ }\href
  {https://doi.org/10.1103/PhysRevA.96.063810} {\bibfield  {journal} {\bibinfo
  {journal} {Phys. Rev. A}\ }\textbf {\bibinfo {volume} {96}},\ \bibinfo
  {pages} {063810} (\bibinfo {year} {2017})}\BibitemShut {NoStop}%
\bibitem [{\citenamefont {Hoang}\ \emph {et~al.}(2016)\citenamefont {Hoang},
  \citenamefont {Ahn}, \citenamefont {Bang},\ and\ \citenamefont
  {Li}}]{NC-7-12550}%
  \BibitemOpen
  \bibfield  {author} {\bibinfo {author} {\bibfnamefont {T.~M.}\ \bibnamefont
  {Hoang}}, \bibinfo {author} {\bibfnamefont {J.}\ \bibnamefont {Ahn}},
  \bibinfo {author} {\bibfnamefont {J.}\ \bibnamefont {Bang}},\ and\
  \bibinfo {author} {\bibfnamefont {T.-C.}\ \bibnamefont {Li}},\ }\bibfield
   {title} {\bibinfo {title} {Electron spin control of optically levitated
  nanodiamonds in vacuum},\ }\href {https://doi.org/10.1038/ncomms12250}
  {\bibfield  {journal} {\bibinfo  {journal} {Nat. Commun.}\ }\textbf {\bibinfo
  {volume} {7}},\ \bibinfo {pages} {12250} (\bibinfo {year}
  {2016})}\BibitemShut {NoStop}%
\bibitem [{\citenamefont {Li}\ \emph {et~al.}(2016)\citenamefont {Li},
  \citenamefont {Xiang}, \citenamefont {Rabl},\ and\ \citenamefont
  {Nori}}]{PRL-112-015502}%
  \BibitemOpen
  \bibfield  {author} {\bibinfo {author} {\bibfnamefont {P.-B.}\ \bibnamefont
  {Li}}, \bibinfo {author} {\bibfnamefont {Z.-L.}\ \bibnamefont {Xiang}},
  \bibinfo {author} {\bibfnamefont {P.}\ \bibnamefont {Rabl}},\ and\
  \bibinfo {author} {\bibfnamefont {F.}\ \bibnamefont {Nori}},\ }\bibfield
  {title} {\bibinfo {title} {Hybrid quantum device with nitrogen-vacancy
  centers in diamond coupled to carbon nanotubes},\ }\href
  {https://doi.org/10.1103/PhysRevLett.117.015502} {\bibfield  {journal}
  {\bibinfo  {journal} {Phys. Rev. Lett.}\ }\textbf {\bibinfo {volume} {117}},\
  \bibinfo {pages} {015502} (\bibinfo {year} {2016})}\BibitemShut {NoStop}%
\bibitem [{\citenamefont {Fornieri}\ \emph {et al.}()\citenamefont {Fornieri},
  \citenamefont {Whiticar}, \citenamefont {Setiawan}, \citenamefont
  {Mar\'{i}n}, \citenamefont {Drachmann}, \citenamefont {Keselman},
  \citenamefont {Gronin}, \citenamefont {Thomas}, \citenamefont {Wang},
  \citenamefont {Kallaher}, \citenamefont {Gardner}, \citenamefont {Berg},
  \citenamefont {Manfra}, \citenamefont {Stern}, \citenamefont {Marcus},\ and\
  \citenamefont {Nichele}}]{arxiv-1089-03037v1}%
  \BibitemOpen
  \bibfield  {author} {\bibinfo {author} {\bibfnamefont {A.}\ \bibnamefont
  {Fornieri}}, \bibinfo {author} {\bibfnamefont {A.~M.}\ \bibnamefont
  {Whiticar}}, \bibinfo {author} {\bibfnamefont {F.}~\bibnamefont {Setiawan}},
  \bibinfo {author} {\bibfnamefont {E.~P.}\ \bibnamefont
  {Mar\'{i}n}}, \bibinfo {author} {\bibfnamefont {A.~C.~C.}\ \bibnamefont
  {Drachmann}}, \bibinfo {author} {\bibfnamefont {A.}\ \bibnamefont
  {Keselman}}, \bibinfo {author} {\bibfnamefont {S.}\ \bibnamefont
  {Gronin}}, \bibinfo {author} {\bibfnamefont {C.}\ \bibnamefont
  {Thomas}}, \bibinfo {author} {\bibfnamefont {T.}\ \bibnamefont {Wang}},
  \bibinfo {author} {\bibfnamefont {R.}\ \bibnamefont {Kallaher}}, \bibinfo
  {author} {\bibfnamefont {G.~C.}\ \bibnamefont {Gardner}}, \bibinfo
  {author} {\bibfnamefont {E.}\ \bibnamefont {Berg}}, \bibinfo {author}
  {\bibfnamefont {M.~J.}\ \bibnamefont {Manfra}}, \bibinfo {author}
  {\bibfnamefont {A.}\ \bibnamefont {Stern}}, \bibinfo {author} {\bibfnamefont
  {C.~M.}\ \bibnamefont {Marcus}}, \ and\ \bibinfo {author} {\bibfnamefont
  {F.}\ \bibnamefont {Nichele}},\ }\bibfield  {title} {\bibinfo
  {title} {Evidence of topological superconductivity in planar Josephson
  junctions},\ }{\bibinfo  {journal} {e-print,}\
  }\href {https://arxiv.org/abs/1809.03037} {\bibfield
  arxiv.org/abs/1809.03037}\BibitemShut {NoStop}%
\bibitem [{\citenamefont {Meyer}\ \emph {et~al.}(2005)\citenamefont {Meyer},
  \citenamefont {Paillet},\ and\ \citenamefont {Roth}}]{SC-309-1539}%
  \BibitemOpen
\bibfield  {journal} {  }\bibfield  {author} {\bibinfo {author} {\bibfnamefont
  {J.~C.}\ \bibnamefont {Meyer}}, \bibinfo {author} {\bibfnamefont
  {M.}\ \bibnamefont {Paillet}},\ and\ \bibinfo {author} {\bibfnamefont
  {S.}\ \bibnamefont {Roth}},\ }\bibfield  {title} {\bibinfo {title}
  {Single-molecule torsional pendulum},\ }\href
  {https://doi.org/10.1126/science.1115067} {\bibfield  {journal} {\bibinfo
  {journal} {Science}\ }\textbf {\bibinfo {volume} {309}},\ \bibinfo {pages}
  {1539} (\bibinfo {year} {2005})}\BibitemShut {NoStop}%
\bibitem [{\citenamefont {Mamin}\ \emph {et~al.}(2007)\citenamefont {Mamin},
  \citenamefont {Poggio}, \citenamefont {Degen},\ and\ \citenamefont
  {Rugar}}]{NatNano-2-301}%
  \BibitemOpen
  \bibfield  {author} {\bibinfo {author} {\bibfnamefont {H.~J.}\ \bibnamefont
  {Mamin}}, \bibinfo {author} {\bibfnamefont {M.}~\bibnamefont {Poggio}},
  \bibinfo {author} {\bibfnamefont {C.~L.}\ \bibnamefont {Degen}},\ and\
  \bibinfo {author} {\bibfnamefont {D.}~\bibnamefont {Rugar}},\ }\bibfield
  {title} {\bibinfo {title} {Nuclear magnetic resonance imaging with 90-nm
  resolution},\ }\href {https://doi.org/10.1038/nnano.2007.105} {\bibfield
  {journal} {\bibinfo  {journal} {Nat. Nanotechnol.}\ }\textbf {\bibinfo
  {volume} {2}},\ \bibinfo {pages} {301} (\bibinfo {year}
  {2007})}\BibitemShut {NoStop}%
\bibitem [{\citenamefont {Romito}\ \emph {et~al.}(2012)\citenamefont {Romito},
  \citenamefont {Alicea}, \citenamefont {Refael},\ and\ \citenamefont {von
  Oppen}}]{PRB-85-020502}%
  \BibitemOpen
  \bibfield  {author} {\bibinfo {author} {\bibfnamefont {A.}\
  \bibnamefont {Romito}}, \bibinfo {author} {\bibfnamefont {J.}\
  \bibnamefont {Alicea}}, \bibinfo {author} {\bibfnamefont {G.}\ \bibnamefont
  {Refael}},\ and\ \bibinfo {author} {\bibfnamefont {F.}\ \bibnamefont {von
  Oppen}},\ }\bibfield  {title} {\bibinfo {title} {Manipulating Majorana
  fermions using supercurrents},\ }\href
  {https://doi.org/10.1103/PhysRevB.85.020502} {\bibfield  {journal} {\bibinfo
  {journal} {Phys. Rev. B}\ }\textbf {\bibinfo {volume} {85}},\ \bibinfo
  {pages} {020502} (\bibinfo {year} {2012})}\BibitemShut {NoStop}%
\bibitem [{\citenamefont {Stanwix}\ \emph {et~al.}(2010)\citenamefont
  {Stanwix}, \citenamefont {Pham}, \citenamefont {Maze}, \citenamefont
  {Le~Sage}, \citenamefont {Yeung}, \citenamefont {Cappellaro}, \citenamefont
  {Hemmer}, \citenamefont {Yacoby}, \citenamefont {Lukin},\ and\ \citenamefont
  {Walsworth}}]{prb-82-201201}%
  \BibitemOpen
  \bibfield  {author} {\bibinfo {author} {\bibfnamefont {P.~L.}\ \bibnamefont
  {Stanwix}}, \bibinfo {author} {\bibfnamefont {L.~M.}\ \bibnamefont {Pham}},
  \bibinfo {author} {\bibfnamefont {J.~R.}\ \bibnamefont {Maze}}, \bibinfo
  {author} {\bibfnamefont {D.}~\bibnamefont {Le~Sage}}, \bibinfo {author}
  {\bibfnamefont {T.~K.}\ \bibnamefont {Yeung}}, \bibinfo {author}
  {\bibfnamefont {P.}~\bibnamefont {Cappellaro}}, \bibinfo {author}
  {\bibfnamefont {P.~R.}\ \bibnamefont {Hemmer}}, \bibinfo {author}
  {\bibfnamefont {A.}~\bibnamefont {Yacoby}}, \bibinfo {author} {\bibfnamefont
  {M.~D.}\ \bibnamefont {Lukin}},\ and\ \bibinfo {author} {\bibfnamefont
  {R.~L.}\ \bibnamefont {Walsworth}},\ }\bibfield  {title} {\bibinfo {title}
  {Coherence of nitrogen-vacancy electronic spin ensembles in diamond},\ }\href
  {https://doi.org/10.1103/PhysRevB.82.201201} {\bibfield  {journal} {\bibinfo
  {journal} {Phys. Rev. B}\ }\textbf {\bibinfo {volume} {82}},\ \bibinfo
  {pages} {201201} (\bibinfo {year} {2010})}\BibitemShut {NoStop}%
\bibitem [{\citenamefont {Tian}\ and\ \citenamefont
  {Wang}(2010)}]{pra-82-053806}%
  \BibitemOpen
  \bibfield  {author} {\bibinfo {author} {\bibfnamefont {L.}~\bibnamefont
  {Tian}}\ and\ \bibinfo {author} {\bibfnamefont {H.-L.}\ \bibnamefont
  {Wang}},\ }\bibfield  {title} {\bibinfo {title} {Optical wavelength
  conversion of quantum states with optomechanics},\ }\href
  {https://doi.org/10.1103/PhysRevA.82.053806} {\bibfield  {journal} {\bibinfo
  {journal} {Phys. Rev. A}\ }\textbf {\bibinfo {volume} {82}},\ \bibinfo
  {pages} {053806} (\bibinfo {year} {2010})}\BibitemShut {NoStop}%
\bibitem [{\citenamefont {Wang}\ and\ \citenamefont
  {Clerk}(2012)}]{PRL-108-153603}%
  \BibitemOpen
  \bibfield  {author} {\bibinfo {author} {\bibfnamefont {Y.-D.}\
  \bibnamefont {Wang}}\ and\ \bibinfo {author} {\bibfnamefont {A.~A.}\
  \bibnamefont {Clerk}},\ }\bibfield  {title} {\bibinfo {title} {Using
  interference for high fidelity quantum state transfer in optomechanics},\
  }\href {https://doi.org/10.1103/PhysRevLett.108.153603} {\bibfield  {journal}
  {\bibinfo  {journal} {Phys. Rev. Lett.}\ }\textbf {\bibinfo {volume} {108}},\
  \bibinfo {pages} {153603} (\bibinfo {year} {2012})}\BibitemShut {NoStop}%
\bibitem [{\citenamefont {Tian}(2012)}]{PRL-108-153604}%
  \BibitemOpen
  \bibfield  {author} {\bibinfo {author} {\bibfnamefont {L.}\ \bibnamefont
  {Tian}},\ }\bibfield  {title} {\bibinfo {title} {Adiabatic state conversion
  and pulse transmission in optomechanical systems},\ }\href
  {https://doi.org/10.1103/PhysRevLett.108.153604} {\bibfield  {journal}
  {\bibinfo  {journal} {Phys. Rev. Lett.}\ }\textbf {\bibinfo {volume} {108}},\
  \bibinfo {pages} {153604} (\bibinfo {year} {2012})}\BibitemShut {NoStop}%
\bibitem [{\citenamefont {Li}\ \emph {et~al.}(2015)\citenamefont {Li},
  \citenamefont {Liu}, \citenamefont {Gao}, \citenamefont {Xiang},
  \citenamefont {Rabl}, \citenamefont {Xiao},\ and\ \citenamefont
  {Li}}]{prapl-04-044003}%
  \BibitemOpen
  \bibfield  {author} {\bibinfo {author} {\bibfnamefont {P.-B.}\ \bibnamefont
  {Li}}, \bibinfo {author} {\bibfnamefont {Y.-C.}\ \bibnamefont {Liu}},
  \bibinfo {author} {\bibfnamefont {S.-Y.}\ \bibnamefont {Gao}}, \bibinfo
  {author} {\bibfnamefont {Z.-L.}\ \bibnamefont {Xiang}}, \bibinfo {author}
  {\bibfnamefont {P.}\ \bibnamefont {Rabl}}, \bibinfo {author}
  {\bibfnamefont {Y.-F.}\ \bibnamefont {Xiao}},\ and\ \bibinfo {author}
  {\bibfnamefont {F.-L.}\ \bibnamefont {Li}},\ }\bibfield  {title} {\bibinfo
  {title} {Hybrid quantum device based on NV centers in diamond
  nanomechanical resonators plus superconducting waveguide cavities},\ }\href
  {https://doi.org/10.1103/PhysRevApplied.4.044003} {\bibfield  {journal}
  {\bibinfo  {journal} {Phys. Rev. Appl.}\ }\textbf {\bibinfo {volume} {4}},\
  \bibinfo {pages} {044003} (\bibinfo {year} {2015})}\BibitemShut {NoStop}%
\bibitem [{\citenamefont {Johansson}\ \emph {et~al.}(2012)\citenamefont
  {Johansson}, \citenamefont {Nation},\ and\ \citenamefont {Nori}}]{CPC}%
  \BibitemOpen
  \bibfield  {author} {\bibinfo {author} {\bibfnamefont {J.~R.}\ \bibnamefont
  {Johansson}}, \bibinfo {author} {\bibfnamefont {P.~D.}\ \bibnamefont
  {Nation}},\ and\ \bibinfo {author} {\bibfnamefont {F.}\ \bibnamefont
  {Nori}},\ }\bibfield  {title} {\bibinfo {title} {Qutip 2: A python framework
  for the dynamics of open quantum systems},\ }\href
  {https://doi.org/10.1016/j.cpc.2012.11.019} {\bibfield  {journal} {\bibinfo
  {journal} {Comput. Phys. Commun.}\ }\textbf {\bibinfo {volume} {184}},\
  \bibinfo {pages} {1234} (\bibinfo {year} {2013})}\BibitemShut {NoStop}%
\end{thebibliography}
%apsrev4-2.bst 2019-01-14 (MD) hand-edited version of apsrev4-1.bst
%Control: key (0)
%Control: author (0) dotless jnrlst
%Control: editor formatted (1) identically to author
%Control: production of article title (0) allowed
%Control: page (1) range
%Control: year (0) verbatim
%Control: production of eprint (0) enabled
\providecommand{\noopsort}[1]{}\providecommand{\singleletter}[1]{#1}%

\end{document}